\documentclass[twocolumn,floatfix,prl]{revtex4-2}%
\usepackage{graphicx}%
\usepackage{amsmath,amssymb}%
\usepackage{amsfonts}
\usepackage{amssymb}
\usepackage{mathrsfs}
\usepackage{color}
\usepackage{bm}
\def\d{{\partial}}
\def\s{{\sigma}}
\def\e{{\epsilon}}
\def\k{{ {\bm k} }}
\def\p{{ {\bm p} }}
\def\q{{ {\bm q} }}

\def\0{{ {\bm 0} }}
\def\w{{\omega}}
\def\a{{\alpha}}
\def\b{{\beta}}

\allowdisplaybreaks[4]

\newcommand{\maru}[1]{\raise0.2ex\hbox{\textcircled{\scriptsize{#1}}}}

\begin{document}
\title{
Odd-parity bond order and induced nonreciprocal transport in the kagome metal CsTi$_3$Bi$_5$ driven by quantum interference
}
\author{
Jianxin Huang$^{1}$, Youichi Yamakawa$^{1}$,
Rina Tazai$^{2}$, Takahiro Morimoto$^{3}$  and Hiroshi Kontani$^1$
}
\date{\today }

\begin{abstract}
Kagome metals present a fascinating platform of quantum phases thanks to the interplay between the geometric frustration and strong electron correlation.
Here, we propose the emergence of the electric odd-parity bond order (BO)
that originates from the intra-unit-cell odd-parity configuration 
in recently discovered kagome metal CsTi$_3$Bi$_5$.
The predicted $E_{1u}$ BO is induced by the beyond-mean-field mechanism, that is, the quantum interference among different sublattice spin fluctuations.
Importantly,
the accompanied nematic deformation of the Fermi surface is just $\sim1$\%
while the intensity of the quasiparticle interference signal
exhibits drastic nematic anisotropy,
consistent with the 
scanning tunneling microscope measurements in CsTi$_3$Bi$_5$.
The present odd-parity BO triggers interesting phenomena,
such as the non-linear Hall effect and emergent electromagnetism.

\end{abstract}

\address{
$^1$Department of Physics, Nagoya University,
Nagoya 464-8602, Japan \\
$^2$Yukawa Institute for Theoretical Physics, Kyoto University,
Kyoto 606-8502, Japan \\
$^3$Department of Applied Physics, The University of Tokyo, Tokyo 113-8656, Japan
}
\sloppy

\maketitle



\subsection{I. INTRODUCTION\vspace{-1em}}
The discovery of kagome metals has greatly enriched the study of condensed matter physics.
The interplay between the geometric frustration and strong electron correlation
gives rise to quantum phases.
For instance, 
$2\times2$ charge-density-wave (CDW) order
\cite{kagome-exp1,kagome-exp2,STM1,STM2},
time-reversal-symmetry (TRS) breaking loop-current order
\cite{muSR3-Cs,muSR2-K,muSR4-Cs,muSR5-Rb,eMChA},
nematic order 
\cite{elastoresistance-kagome,birefringence-kagome,STM2},
and superconductivity,
\cite{Roppongi,SC2},
have been discovered in the V-based kagome metal $A$V$_3$Sb$_5$
($A$= Cs, Rb, K).
Similar quantum phase transitions 
(such as the $\sqrt{3}\times\sqrt{3}$ CDW without TRS)
are observed in bilayer kagome metal ScV$_6$Sn$_6$
\cite{arXiv:2304.06436}.
Various theoretical studies have been conducted on the origin of 
quantum states in kagome metals
\cite{Thomale2021,Neupert2021,Balents2021,Tazai-kagome,Tazai-kagome2,Tazai-kagome-GL,Fernandes-GL,Thomale-GL,Nat-g-ology,
fRG for vHS,fRG for nem,Ti-ARPES}, 
by focusing on the strong correlation and geometric frustration.
However, numerous essential electronic properties remain unresolved.

The recent discovery of a Ti-based kagome superconductor (SC) CsTi$_3$Bi$_5$
\cite{CsTiBi5-first,arXiv:2209.11656,arXiv:2211.16477,arXiv:2211.12264,ARPES-theory,CsTiBi5-ARPES,CsTiBi5-transport,Ti-ARPES,CsTiBi5-oscillation}
has revealed that further exotic electronic states emerge.
While no CDW occurs that breaks translational symmetry, 
CsTi$_3$Bi$_5$ exhibits quantum phases similar to V-based kagome metals, 
such as nematicity and superconductivity ($T_{\rm c}=4.8$K).
Nematic order has been revealed by 
scanning tunneling microscope (STM) measurements
\cite{arXiv:2211.16477,arXiv:2211.12264},
and its transition temperature is $T_{0}\sim100$K
according to angular-dependent magnetoresistance 
\cite{arXiv:2211.12264}.
The wave vector of the order parameter is $\q={\bm0}$
because of no Fermi surface (FS) reconstruction
\cite{arXiv:2211.16477,arXiv:2211.12264}.
Notably, however, the nematicity in CsTi$_3$Bi$_5$
has characteristic properties that would be 
distinct from other nematic metals.
Also, the lattice deformation and the kink in the resistivity 
at $T\sim T_0$ are almost invisible.
For $T\ll T_0$, in contrast, small nematic deformation of the FS 
leads to drastic nematicity in the quasiparticle (QP) scattering\cite{arXiv:2211.16477,arXiv:2211.12264} and the angle-resolved photoemission spectroscopy (ARPES) spectrum \cite{ARPES-theory}.
These facts indicate the emergence of a quantum state
in Ti-based kagome metals.




Importantly, electronic nematic order ($\q={\bm0}$)
transcends the realm of mean-field (or classical) order,
where FS nesting ($\q\ne{\bm0}$) leads to kinetic energy gain.
That is, nematicity without band folding is a 
hallmark of nontrivial quantum correlations.
A famous example is Fe-based SCs,
where nematic order with orbital polarization
is caused by beyond-mean-field electron correlations
\cite{Onari-SCVC,Yamakawa-FeSe,Onari-TBG,Tsuchiizu1,fRG-sin-orb,Tsuchiizu4,fRG-BEDT,Tazai-rev2021,Chubukov-PRX2016,Fernandes-rev2018,Kontani-AdvPhys,Davis-rev2013}.
In the FeSe family, the nematic quantum critical point (QCP) gives the critical behaviors and pairing mechanism
\cite{Shibauchi-FeSeTe}.
In contrast, in Ti-based kagome metals,
on-site orbital degeneracy is absent,
and induced anomalies in the resistivity and lattice constant
are quite small.
Therefore, the origin of unusual nematicity in a Ti-based kagome metal
and its relation to other nematic metals are highly nontrivial.

\begin{figure}[htb]
\includegraphics[width=.99\linewidth]{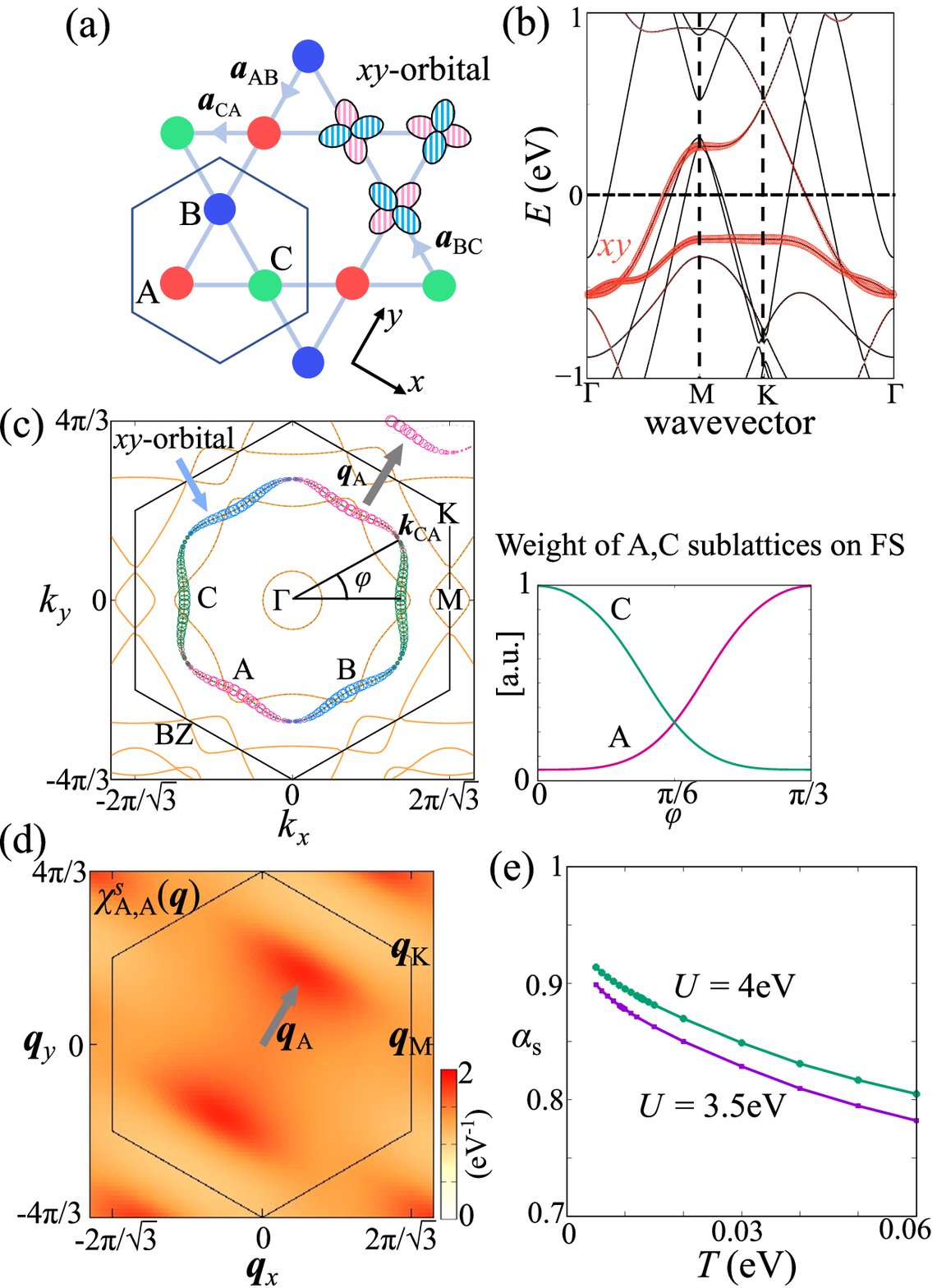}
\caption{
(a) Kagome lattice structure composed of Ti ions.
The unit cell contains three sublattices A (red), B (blue), and C (green).
(b) Band structure of CsTi$_3$Bi$_5$ model.
(c) Fermi surfaces (FSs) of CsTi$_3$Bi$_5$ model.
The $xy$-orbital weights of A, B, and C sublattices are depicted 
by red, blue and green colors, respectively.
The sublattice density of states (DOS) on the $xy$-orbital FS is shown.
(d) Spin susceptibility $\chi^s_{\rm A,A}(\q)$
for $U=4$eV at $T=0.01$eV.
(e) Stoner factor $\a_S$ for $U=4$eV and $3.5$eV as function of $T$.
}
\label{fig:fig1}
\end{figure}

In this paper, we find that 
the $E_{1u}$ symmetry bond order (BO) is induced by the 
intersublattice attraction due to the paramagnon interference mechanism.
Here, intra-unit-cell staggered BO
leads to the nonpolar odd-parity state,
which has rarely been studied in strongly correlated metals.
The $E_{1u}$ BO explains the almost invisible
anomalies in the resistivity and lattice constant at $T\lesssim T_0$
because $\Delta k_F \propto \phi^2 \ (\propto T_0-T)$.
For $T\ll T_0$, however, large $E_{1u}$ BO ($\phi\gg T_0$)
causes remarkable nematicity in the QP interference (QPI) signal \cite{arXiv:2211.16477,arXiv:2211.12264} 
and ARPES spectrum \cite{ARPES-theory}
observed in CsTi$_3$Bi$_5$.
Interestingly, we reveal that the odd-parity $E_{1u}$ BO triggers the nonreciprocal nonlinear Hall (NLH) effect.


\subsection{II. MODEL HAMILTONIAN\vspace{-1em}}
\begin{figure}[htb]
\includegraphics[width=.99\linewidth]{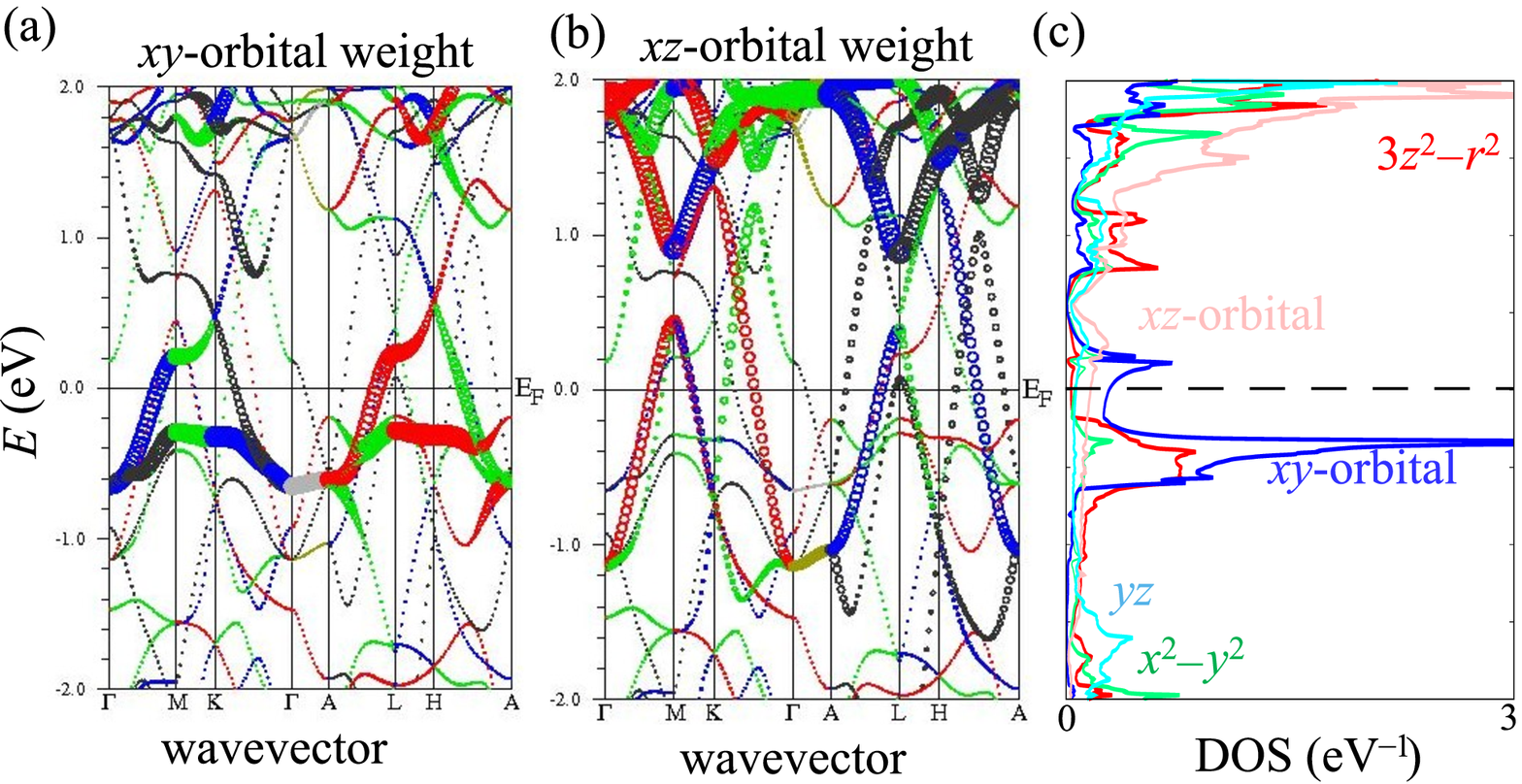}
\caption{
Band structures with (a) $xy$-orbital weight 
and (b) $xz$-orbital weight denoted by the size of each circle.
(c) Each $d$-orbital density of states (DOS).
}
\label{fig:fig7}
\end{figure}

The two-dimensional (2D) kagome lattice structure of 
CsTi$_3$Bi$_5$ is shown in Fig. \ref{fig:fig1}(a).
Each unit cell is composed of three Ti-ion sublattices A, B and C.
We derive the 30 orbital tight-binding model with 
15 Ti $d$ orbitals and 15 Bi $p$ orbitals
based on the band structure given by {\footnotesize WIEN}2{\footnotesize K} software,
which is shown in Appendix A.
The $d$-electron FSs are mainly composed of the
$xy$ orbital in addition to the $xz$ orbital
[$N_{xz}(0)\sim0.4 N_{xy}(0)$ as shown in Fig. \ref{fig:fig7}(c)].
The number of electrons in a Ti-based system 
per Ti ion is one less than in the V-based system, 
leading to smaller FSs with different $d$-orbital character.
Notably, the Van Hove singularity (VHS) points, 
which play an important role in V-based systems, 
are far away from the Fermi level in the Ti-based system. 
The band structure of the 30 orbital model
is shown in Fig. \ref{fig:fig1}(b).


Figure \ref{fig:fig1}(c) exhibits the FS of the present 2D model,
which reproduces the ARPES measurement well
\cite{Ti-ARPES}.
The $xy$-orbital weights of A, B, and C sublattices are 
depicted by red, blue, and green colors, respectively.
[The inset shows the sublattice density of states (DOS) on the $xy$-orbital FS.]
There is a prominent intrasublattice nesting at $\q=\q_{\rm A}$,
in high contrast with the absence of the intrasublattice nesting 
in the pure-type FS of V-based kagome metals 
(called {\it sublattice interference})
\cite{Thomale2021,Tazai-kagome}.
The DOS at the Fermi level is mainly composed of
the $xy$ orbital, and we verified that the 
spin fluctuations develop only in the $xy$ orbital
based on the multiorbital random-phase approximation (RPA),
as we explain in Appendix A.
Therefore, we introduce the Coulomb interaction only on the $xy$ orbital
in this paper.
Hereafter, the unit of the energy is eV unless otherwise noted.

\subsection{III. SPIN SUSCEPTIBILITY AND SELF-ENERGY\vspace{-1em}}
Here, we calculate the spin susceptibility $\chi^s_{l,m}(\q)$ self-consistently
by including the spin-fluctuation-induced self-energy $\Sigma_{l,m}(k)$
of the $xy$-orbital electrons.
Here, $k\equiv[\k,\e_n=(2n+1)\pi T]$ and $l,m={\rm A, B, C}$.
We use the fluctuation-exchange (FLEX) approximation
\cite{Bickers2,Kontani-RH,Kontani-rev1},
which is explained in Appendix B.
The obtained spin susceptibility $\chi^s_{\rm A,A}(\q)$
for $U=4$eV at $T=0.01$eV is shown in Fig. \ref{fig:fig1}(d). 
Due to the geometrical frustration,
$\chi^s_{\rm A,A}(\q)$ exhibits a very broad peak around the 
nesting vector $\q_{\rm A}$ in Fig. \ref{fig:fig1}(c),
which is favorable for the paramagnon interference mechanism
given by the convolution of two $\chi^s$'s [see Fig. \ref{fig:fig2}(a)].
Note that $\chi^s_{l,m}(\q)$ is small for $l\ne m$,
meaning that the spin susceptibility is sublattice selective.
Figure \ref{fig:fig1}(e) shows the Stoner factor 
$\a_S\equiv\max_\q U\chi^0(\q)$.
Magnetism appears when $\a_S\ge1$.
Thus, the system remains paramagnetic until low temperatures 
owing to the geometrical frustration.

\subsection{IV. LINEARIZED DENSITY-WAVE EQUATION\vspace{-1em}}
The nonlocal nature of the BO states is not obtained in FLEX approximation.
However, the BO can be induced by the beyond-FLEX nonlocal correlations,
called {\it vertex corrections} (VCs)
\cite{Onari-SCVC,Yamakawa-FeSe,Onari-TBG,Kontani-AdvPhys}.
The BOs due to VCs are derived from the linearized density-wave (DW) equation
\cite{Kontani-AdvPhys}
\begin{eqnarray}
\lambda_{\q}f_\q^{L}(k)&=& -\frac{T}{N}\sum_{p,M_1,M_2}
I_\q^{L,M_1}(k,p) 
\nonumber \\
& &\times \{ G(p)G(p+\q) \}^{M_1,M_2} f_\q^{M_2}(p) ,
\label{eqn:DWeq}
\end{eqnarray}
where $L\equiv (l,l')$ and $M_i\equiv(m_i,m_i')$ 
represent the pair of sublattices A, B, and C.
Here, $I_\q^{L,M}(k,p)$ is the electron-hole pairing interaction.
It is uniquely derived from the functional derivative of the 
FLEX self-energy to satisfy the conserving laws
$I_{q=0}^{L,M}(k,p)=\delta \Sigma_L(k)/\delta G_M(p)$
\cite{Tazai-LW}.
Here, $\lambda_{\q}$ is the eigenvalue that represents the
instability of the DW at wavevector $\q$, and
$\max_\q\{\lambda_\q\}=1$ at $T=T_0$.
Hereafter, the form factor is normalized as
$\max_{l,m,\k}|f_\q^{l,m}(\k)|=1$.
The physical meaning of the form factor and
expression of the kernel function $I_\q^{L,M}(k,p)$
are given in Appendixes C and D, respectively.

\begin{figure}[htb]
\includegraphics[width=.99\linewidth]{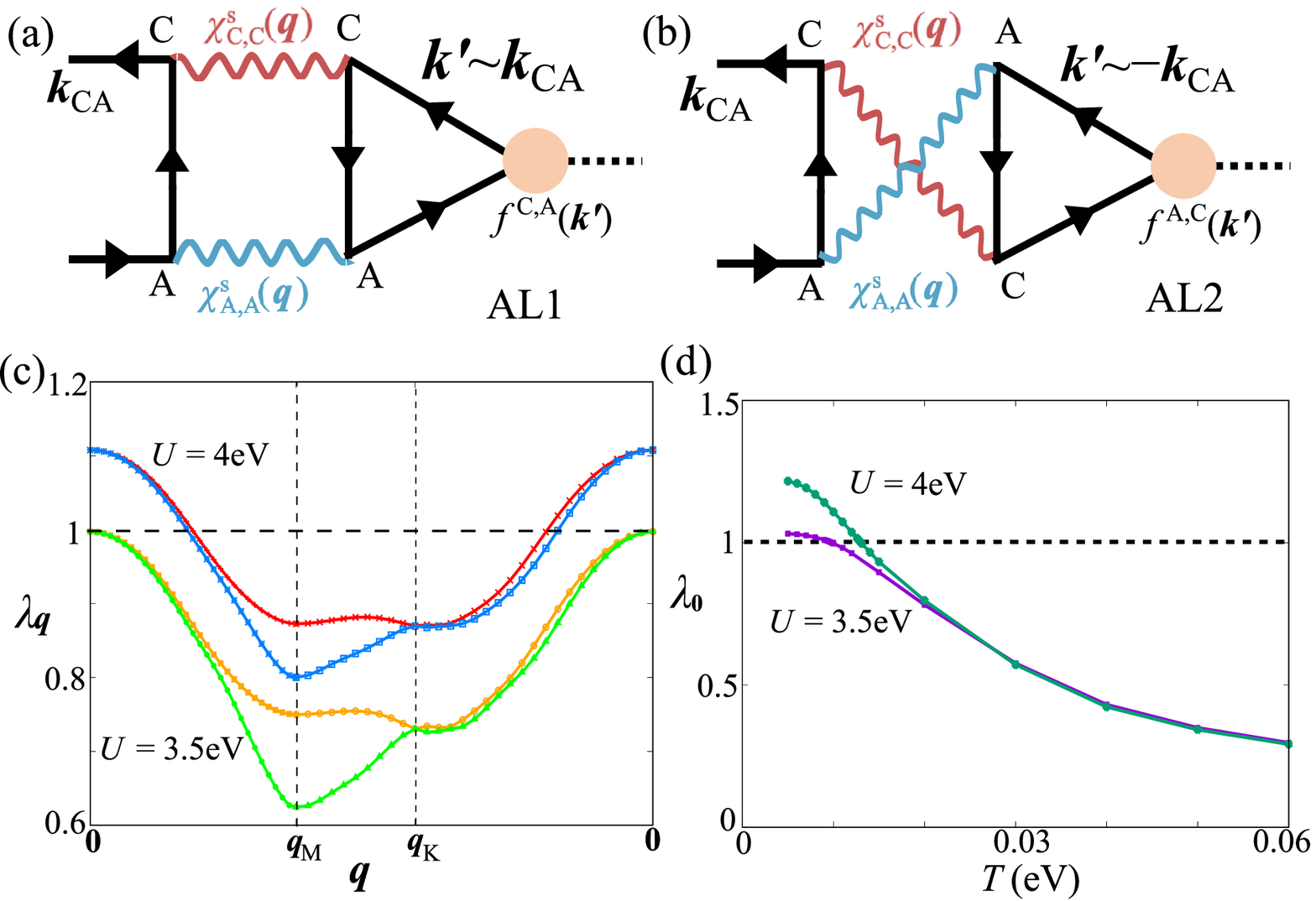}
\caption{
(a) AL1 and 
(b) AL2 processes that give the bond order (BO) at $\q_{\rm BO}={\bm0}$.
Note that $f^{\rm{A,C}}(-\k)=-f^{\rm{A,C}}(\k)=f^{\rm{C,A}}(\k)$ in the $E_{1u}$ state.
(c) $\q$ dependence of the largest and the second largest eigenvalues for $U=4$ and 3.5.
$\q_{\rm M}$ and $\q_{\rm K}$ are shown in Fig. \ref{fig:fig1}(d).
The horizontal line $\lambda=1$ represents the instability of the density-wave (DW) order. Here, the $\lambda_{\q}$ maximum value at $\q=\0$ is $>1$ in both cases. Therefore, the $\q=\0$ DW order is realized.
(d) $T$ dependence of the largest $\lambda_\q$ at $\q={\bm0}$.}
\label{fig:fig2}
\end{figure}

Here, $f_\q^L(k)$ is the Hermitian form factor that is proportional to
particle-hole (p-h) condensation 
$\sum_\s \{ \langle c_{\k+\q,l,\s}^\dagger c_{\k,l',\s}\rangle - \langle \cdots \rangle_0 \}$
or, equivalently, the symmetry-breaking component in the self-energy. 
The kernel function $I$ in Eq. (\ref{eqn:DWeq}) contains the 
Aslamazov-Larkin (AL) terms shown in Figs. \ref{fig:fig2}(a) and \ref{fig:fig2}(b),
which originate from the spin-fluctuation-induced reduction in the free energy promoted by the BO.
Importantly, the interference between two paramagnons gives rise to
the charge-channel DW at $\q_{\rm BO}={\bm0}$ constructively; 
see Appendix B for detail.
In FeSe, the AL terms drive nonmagnetic nematic order 
\cite{Onari-SCVC}.
The importance of AL terms was
verified by previous comparison research between the DW equation and functional renormalization group (fRG) study, which includes higher-order VCs with Maki-Thompson (MT) and AL terms
\cite{Tazai-rev2021,Tsuchiizu1,Tsuchiizu4,fRG-BEDT,fRG-sin-orb}.
Both the fRG study and the DW equation will derive the same order (see Appendix D).

It is notable that Figs. \ref{fig:fig2}(a) and \ref{fig:fig2}(b) only show the contribution of the spin fluctuations. In the DW equation calculation, both spin fluctuations and charge fluctuations have been considered (see Appendix D). However, the contribution of the charge fluctuations is significantly smaller than that of the spin fluctuations. Therefore, here, we only discuss the spin fluctuations.

Here, we solve the charge-channel DW Eq.(\ref{eqn:DWeq}).
As shown in Fig. \ref{fig:fig2}(c),
the eigenvalue takes the maximum at $\q={\bm0}$
in a Ti-based kagome metal for both $U=4$eV and $3.5$eV at $T=0.01$eV.
As we will explain below, the obtained doubly degenerate eigenvalues 
at $\q={\bm0}$ give the odd-parity $E_{1u}$ BO,
which gives nematic FS deformation, as shown in Appendix E.
The $T$ dependence of $\lambda_{\q={\bm0}}$
is shown in Fig. \ref{fig:fig2}(d).
The transition temperature is $T_0\approx0.013$eV for $U=4$eV.
Note that the even-parity $E_{2g}$ symmetry BO
gives the second largest instability;
see Appendix F.


Next, we derive the full DW equation in the FLEX scheme
by following Ref. \cite{Tazai-LW}, which will give the order parameter under the $E_{1u}$ BO transition temperature.
The total self-energy is given as
\begin{eqnarray}
    {\hat \Sigma}(k)={\hat \Sigma}^0(k)+\delta {\hat t}(k) ,
    \label{eqn:self-total}
\end{eqnarray}
where $\Sigma^0$ is the normal self-energy without any symmetry breaking
given by FLEX (see Appendix B)
(Here, we calculate $\Sigma^0$ at each $T$
by subtracting its static and Hermitian part, 
$\Sigma^{0,{\rm H}}(\k)\equiv[\Sigma^0(\k,+i\delta)+\Sigma^0(\k,-i\delta)]/2$,
in order to fix the shape of the FS.)
Next, we derive the symmetry breaking part $\delta t$ self-consistently
based on the following procedure: 
(a) We first calculate 
$\displaystyle S_k\equiv \frac{T}{N}\sum_q G_{k+q}[\Sigma] V_q[\Sigma]$,
where $G_k[\Sigma]$ and $V_q[\Sigma]$ are functions of the total self-energy.
(b) Next, we derive $\delta t$ as
\begin{eqnarray}
\delta t_k =(1-P_{0})S_k ,
\label{eqn:full-DWeq-pro}
\end{eqnarray}
where $P_{0}$ is the projection operator for 
the totally symmetric ($A_{1g}$) channel.
(c) The total self-energy is given as $\Sigma=\Sigma_0+\delta t$.
We repeat (a)$-$(c) until $\delta t$ converges.
It is easy to show that the 
full DW equation is equivalent to the linearized DW equation 
when $\delta t$ is very small.

In Ref. \cite{Tazai-LW},
the authors performed the full DW equation analysis for FeSe, 
which is a typical Fe-based SC.
Electronic nematic order without magnetization 
and its typical size in FeSe ($\phi\sim 50$meV at $T\sim0$)
are satisfactorily obtained.

\begin{figure}[htb]
\includegraphics[width=.99\linewidth]{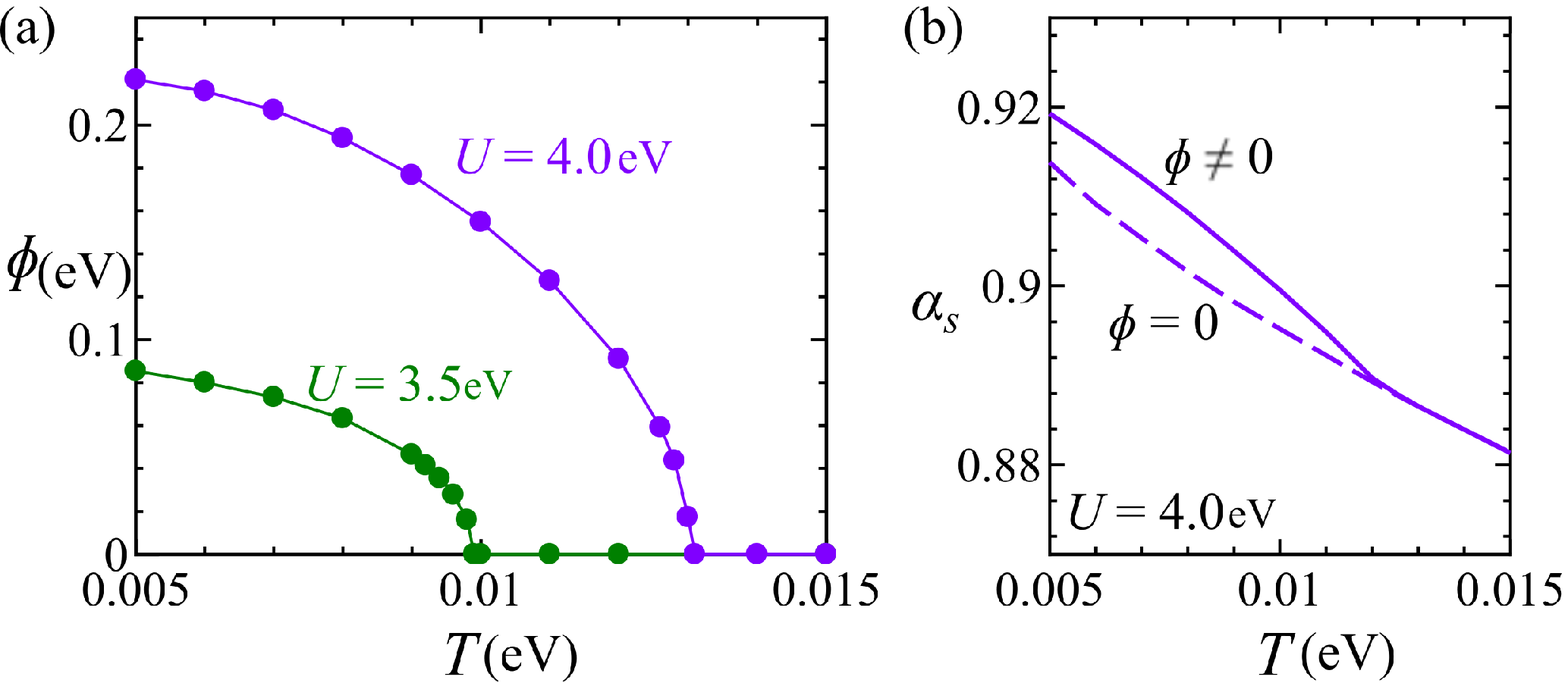}
\caption{
(a) Obtained $E_{1u}$-symmetry bond order (BO) $\phi$
derived from the full density-wave (DW) equation at $U=4$eV ($U=3.5$eV).
The second-order transition occurs at $T_0\approx 13$meV ($T_0\approx 10$meV),
which is consistent with the linearized DW equation analysis 
in Fig. \ref{fig:fig2}(d).
(b) Obtained spin Stoner factor $\a_S$
when $\phi=0$ and $\phi\ne0$ at $U=4$eV.
The increment of $\a_S$ in the $E_{1u}$-symmetry BO state
gives rise to the free-energy gain
through the Luttinger-Ward (LW) function $\Phi$; see Appendix D.
}
\label{fig:full-DW}
\end{figure}

From now on, 
we perform the full DW equation analysis for a Ti-based kagome metal.
In the numerical study, 
we assume that $\delta t(\k)$ is given as $\phi f(\k)$,
where $f(\k)$ is the $E_{1u}$ BO form factor given by the DW equation
and $\phi$ is a constant.
This assumption is well satisfied for nematic order 
in Fe-based SC FeSe \cite{Tazai-LW}.
Under this simplification, we have only to obtain the constant $\phi$
self-consistently numerically.
Figure \ref{fig:full-DW}(a) represents the 
obtained $E_{1u}$ BO parameter $\phi$.
(Here, $z\equiv [1-\d\Sigma_\k(\e)/\d\e|_{\e=0}]^{-1} \approx 0.25$ on the FS.)
Importantly, the second-order transition occurs at $T_0\approx 13$meV,
which is consistent with the linearized DW equation analysis 
in Fig. \ref{fig:fig2}(d).
In Fig. \ref{fig:full-DW}(a), $\phi\sim0.2$eV for $U=4$eV at $T\ll T_0$.
The obtained ratio $z\phi/T_0\approx4$.
This is larger than the BCS ratio ($\sim2$) because $\q={\bm0}$ BO does not cause the gap in the DOS so the negative feedback is small
\cite{Tazai-LW}.

Thus, we can set $\phi\sim0.2$eV in analyzing the nematic QPI signal 
in CsTi$_3$Bi$_5$ ($T_0\sim 100$K) in the following discussions.
[The renormalized order parameter observed by ARPES is $\phi^*\equiv z\phi$,
while the band dispersion is also renormalized by $z$.
Therefore, $\phi$ should be used to analyze the FS deformation 
based on the original (bare) band structure.]

Figure \ref{fig:full-DW}(b) shows the spin Stoner factor 
$\a_S$ as a function of $T$ for $\phi=0$ and $\phi\ne0$ ($E_{1u}$ BO) 
derived from ${\hat G}^\phi(k)$, the Green's function in the ordered state.
The obtained $\a_S$ increases in the $E_{1u}$-symmetry BO state.
This result indicates the free-energy gain due to the $E_{1u}$ BO 
because the Luttinger-Ward (LW) function $\Phi$ is reduced as $\a_S\rightarrow1$.
(Note that $\Phi$ represents the reduction of the free energy 
due to the bosonic fluctuations.)

\subsection{V. ODD-PARITY $E_{1u}$ BO SOLUTION\vspace{-1em}}
\begin{figure}[htb]
\includegraphics[width=.99\linewidth]{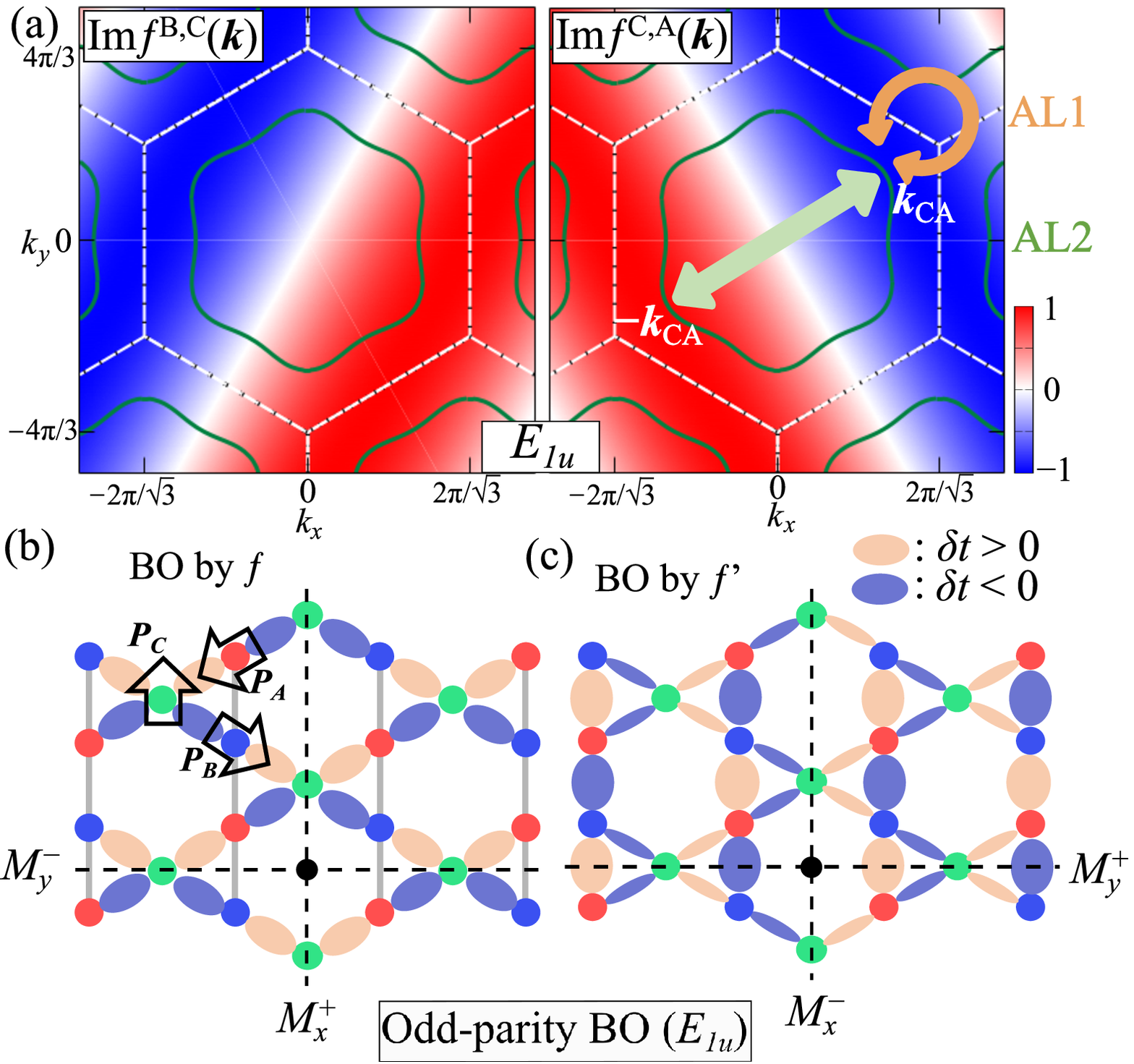}
\caption{
(a) Odd-parity form factors Im$f^{\rm B,C}(\k)$ and Im$f^{\rm C,A}(\k)$ 
at wave vector $\q={\bm0}$.
(Note that Im$f^{\rm B,C}(\k)\propto i\sin \k\cdot{\bm a}_{\rm BC}$
is periodic in the extended Brillouin zone (BZ).)
The original $xy$-orbital Fermi surface (FS) is shown in each panel.
(b) $E_{1u}$ bond order (BO = modulation of the hopping integrals) 
in real space derived from the form factor ${\hat f}(\k)$ in (a).
The parity of $M_{x(y)}$ operation is shown by its superscript ($\pm$).
The electric dipole at each sublattice ${\rm{{\bm P}}}_l$ cancels in total.
(c) $E_{1u}$ BO in real space derived from ${\hat f}'$,
which is orthogonal to ${\hat f}$.
}
\label{fig:fig4}
\end{figure}

Here, we analyze the symmetry of the obtained form factor.
Figure \ref{fig:fig4}(a) exhibits one of the 
doubly degenerate form factors at $\q={\bm0}$,
$f^{\rm B,C}(\k)$ and $f^{\rm C,A}(\k)$, for $U=4$eV at $T=0.01$eV. 
The obtained $f^{l,m}(\k)$ is pure imaginary and odd party
with respect to $\k\rightarrow-\k$ [$f^{l,m}(\k)=-f^{l,m}(-\k)$]
and $l\leftrightarrow m$ [$f^{l,m}(\k)=-f^{m,l}(\k)$].
Note that $f_{\rm CA}(\k)\propto i\sin \k\cdot{\bm a}_{\rm CA}$.
Its BO in real space is
$\delta t_{i,j}\propto \sum_\k f^{l,m}(\k){\rm{exp}}[i\k\cdot({\bm r}_{i}-{\bm r}_{j})]$, 
which is real and even parity $\delta t_{i,j}=\delta t_{j,i}$.
Note that $i\ (j)$ is the site index of sublattice $l\ (m)$.
Figure \ref{fig:fig4}(b) depicts the BO in real space
derived from the Fourier transform of the form factor in 
Fig. \ref{fig:fig4}(a).
Its orthogonal state is shown in Fig. \ref{fig:fig4}(c).
The parity of the mirror operation $M_{x(y)}$, 
$x(y)\rightarrow -x(-y)$, is shown by its superscript ($\pm$)
in Figs. \ref{fig:fig4}(b) and \ref{fig:fig4}(c).
Thus, the parity of the inversion is $I=M_x M_y=-1$ (odd parity).
The electric field gradient at each Ti site induced by the BO results in electric dipole order. 
Since the electric dipole at each sublattice cancels in total [as shown in Fig. \ref{fig:fig4}(b)],
this is not a ferroelectric metal.
Apparently, the nonpolar odd parity originates from the
intra-unit-cell staggered BO.
It can be called the electric toroidal quadrupole BO by focusing on the dipole moments denoted as $P_A$, $P_B$, and $P_C$. It may also be interpreted as electric octupole BO.
Such odd-parity BO 
is rarely studied in strongly correlated metals.
Note that the electric toroidal quadrupole state in the pyrochlore metal Cd$_2$Re$_2$O$_7$
\cite{Cd2Re2O7-1,Cd2Re2O7-2,Cd2Re2O7-3,SHG-Cd2Re2O7}
is closely tied to the strong spin-orbit coupling (SOC),
while SOC is unnecessary in the present mechanism for a Ti-kagome metal.


Now we explain why $E_{1u}$ BO is caused by the AL terms
in Figs. \ref{fig:fig2}(a) and \ref{fig:fig2}(b), which give
the left-hand side of the DW equation $\lambda_{\q={\bm0}} f^{\rm C,A}(\k)$.
In this model, $\chi_{l,m}^s$'s are large for $l=m$ 
({i.e.}, sublattice selective),
and the AL1 term in Fig. \ref{fig:fig2}(a) [AL2 term in Fig. \ref{fig:fig2}(b)]
gives the attraction between $\k$ and $\k'\approx \k$
($\k'\approx -\k$)
due to the particle-particle (p-h) process included in AL1 [AL2] 
\cite{Onari-b2g}.
By setting $\k=\k_{\rm CA}$ shown in Fig. \ref{fig:fig1}(c), we obtain 
\begin{eqnarray}
&&{\rm [AL1]^{C,A}}(\k_{\rm CA})\sim I_{\rm AL} N(0) f^{\rm C,A}(\k_{\rm CA}),
\\
&&{\rm [AL2]^{C,A}}(\k_{\rm CA})\sim I_{\rm AL} N(0) f^{\rm A,C}(-\k_{\rm CA}),
\end{eqnarray}
where $I_{\rm AL}>0$ is the attraction,
and $N(0)$ is the $xy$-orbital DOS per sublattice.
Because $f^{\rm A,C}(-\k_{\rm CA})=-\{f^{\rm C,A}(\k_{\rm CA})\}^*=f^{\rm C,A}(\k_{\rm CA})$,
[AL1] and [AL2] cooperatively contribute to the 
odd-parity BO shown in Fig. \ref{fig:fig4}(a).
[Here, the sublattice degrees of freedom are essential
because AL1 and AL2 cancel for the 
intrasublattice odd-parity BO; $f^{l,l}(\k)=-f^{l,l}(-\k)$.]
The attractions work between the two points of 
the orange (by AL1) and green (by AL2) arrows.
Note that
the intersublattice form factor $f^{l,m}(\k)$ ($l\ne m$)
is not periodic in the first Brillouin zone (BZ)
due to the extra phase factor ${\rm{exp}}[i\k\cdot({\bm r}_l-{\bm r}_m)]$.
The $E_{1u}$ BOs in real space are shown in 
Figs. \ref{fig:fig4}(b) and \ref{fig:fig4}(c).
Each $E_{1u}$ BO changes its sign by the inversion operation.

Interestingly, the quantum interference mechanism
explains both the odd-parity BO in a Ti-based kagome metal
and the $2\times2$ even-parity BO in a V-based one
\cite{Tazai-kagome} on the same footing; 
see Appendix H.
The present electron-correlation mechanism is distinguishable 
from the electron-lattice coupling mechanism of polarmetal transition
that accompanies large lattice distortion
\cite{polar-metal}. In this case, the accompanied lattice deformation is tiny in general.


\subsection{VI. NEMATICITY OF PHYSICAL QUANTITIES\vspace{-1em}}
\begin{figure}[htb]
\includegraphics[width=.99\linewidth]{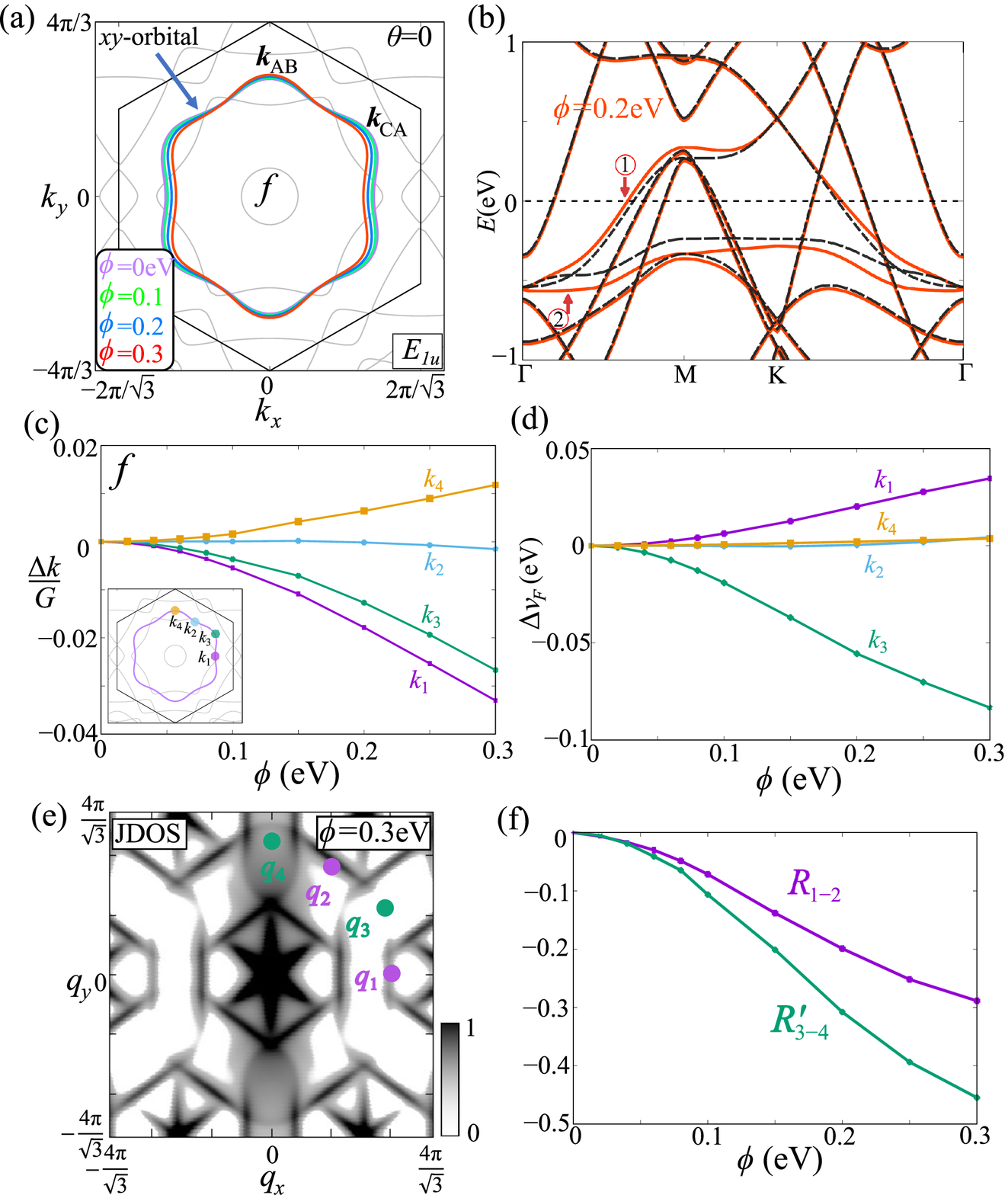}
\caption{
(a) Nematic Fermi surface (FS) deformation due to the $E_{1u}$-symmetry self-energy
$\delta{\hat t}_\k^{\theta=0}=\phi {\hat f}_{\theta=0}(\k)$
for $\phi=0-0.2$eV.
(b) Band dispersion deformation at $\phi=0$ and $0.2$eV.
(c) Deviation of the Fermi momentum $\Delta k_F/G$, where 
$G=4\pi/\sqrt{3}$. The inset shows the Fermi momentums $\k_i$.
(d) Deviation of the Fermi velocity $\Delta v_F$.
(e) Normalized nematic quasiparticle interference (QPI) signal induced by the $E_{1u}$ bond order (BO)
$\delta{\hat t}_\k^{\theta=0}=\phi {\hat f}_{\theta=0}(\k)$ for $\phi=0.3$eV.
$\q_1\sim\q_4$ are the typical QPI momenta shown in (e).
Note that $\q_i\approx 2\k_i$, where 
$\k_i$ is the Fermi momentum shown in the inset of (c).
(f) Anisotropy of the QPI intensity $R_{1-2}\equiv(I_1^\phi-I_2^\phi)/(I_1^\phi+I_2^\phi)$
and $R_{3-4}'\equiv(I_3^\phi-I_4^\phi)/(I_3^\phi+I_4^\phi)$ as a function of $\phi$.
Because $R_{1-2}$ and $R_{3-4}'$ are negative,
the QPI signal is smaller (larger) at $\q=\q_1,\q_3$ ($\q=\q_2,\q_4$).
(Note that $R_{2-1} =-R_{1-2}$ and $R_{3-4}'=-R_{4-3}'$.)
Thus, the QPI signal due to the $E_{1u}$ BO
exhibits sizable nematic anisotropy that is comparable with experimental results.
}
\label{fig:fig5}
\end{figure}

Since the solution for $\q={\bm0}$ is doubly degenerate, 
there exists another form factor ${\hat f}'$ orthogonal to ${\hat f}$, 
and $(f,f')$ belongs to the $E_{1u}$ representation.
Each ${\hat f}$ and ${\hat f}'$ satisfies the
Hermitian condition $f^{lm}(\k)=\{f^{ml}(\k)\}^*$.
Then any linear combination 
${\hat f}_\theta\equiv {\hat f} \cos\theta + {\hat f}' \sin\theta$
gives the solution of Eq. (\ref{eqn:DWeq})
without changing the eigenvalue.
[The coefficients should be real to satisfy
the Hermitian condition of ${\hat f}_\theta(\k)$;
see Appendix C.]
To see the FS deformation, 
we introduce the symmetry-breaking self-energy 
due to the BO state as
$\delta{\hat t}_\k^\theta=\phi {\hat f}_\theta(\k)$
($\max_{l,m,\k}|\delta t^{l,m}_\k|\approx\phi$).
The FSs derived from the eigenvalues of 
${\hat h}_\k^0+\delta{\hat t}_\k^{\theta=0}$,
where ${\hat h}_\k^0$ is the tight-binding model,
for $\phi=0-0.3$eV, are shown in Fig. \ref{fig:fig4}(a).
[The director of the nematic FS is parallel to $(\cos\theta,\sin\theta)$;
see Appendix E.]
The FS deformation due to the $E_{1u}$ BO is 
tiny for $\phi\lesssim0.1$ because it is proportional to $\phi^2$,
while it becomes comparable with experimental nematicity
as shown in Fig. \ref{fig:fig5}(a).


The band dispersion for $\phi=0$ and $0.2$eV are shown in Fig. \ref{fig:fig5}(b).
Large band splitting appears around \maru{2},
while the band shift around the Fermi level \maru{1} is relatively small. 
The maximum change in the Fermi momentum $\Delta k_F/G$
and that in the Fermi velocity $\Delta v_F$ 
are shown in Figs. \ref{fig:fig5}(c) and \ref{fig:fig5}(d), respectively.
They are proportional to $\phi^2$ in the $E_{1u}$ BO,
and the $\k$ points are shown in Fig. \ref{fig:fig5}(a).
Here, $G=4\pi/\sqrt{3}$ is the reciprocal lattice constant,
and the averaged Fermi velocity is $v_F\sim0.5$.
Therefore, the relation $|\Delta k_F/G| \ll |\Delta v_F/v_F|$ holds.

Additionally, it is notable that the intra-unit-cell order ($\q=\0$) does not induce the pseudogap due to the band hybridization \cite{Nem-R-1,Nem-R-2}. In addition, the FS nematic deformation due to the $E_{1u}$ BO is proportional to $(T_0-T)^{3/2}$, where $T_0$ is the transition temperature. Therefore, the proposed $\q=\0$ $E_{1u}$ BO will induce very tiny 
anomalies in the resistivity and thermodynamic quantities at $T=T_0$, consistent with the experiments.
Moreover, by comparing with the even-parity $E_{2g}$ BO (see Appendix F), the odd-parity $E_{1u}$ BO exhibits smaller velocity deviation and FS deformation as $\Delta k \approx \phi^2$, shown in Figs. \ref{fig:fig5}(c) and \ref{fig:fig5}(d).
Therefore, our proposed $\q=\0$ odd-parity BO is more consistent with the experiments than the even-parity BO.

Then we simulate the anisotropic QPI signal of the $E_{1u}$ BO. In the $E_{1u}$ BO state,
the elastic scattering also becomes anisotropic.
The impurity scattering strength with the wave vector $\q$ at energy $E$ is
$n_{\rm imp}{\rm Im}\sum_{\k,l}{\hat G}^\phi(\k,E){\hat T}_l{\hat G}^\phi(\k+\q,E)$,
where $n_{\rm imp}$ is the impurity concentration
and ${\hat T}_l$ is the $T$ matrix due to a single impurity at sublattice $l$. 
Here, we consider the Ti-site unitary impurity potential
represented as $({\hat T}_l)_{m,m'}\sim [-i/\pi N(0)]\delta_{l,m}\delta_{m,m'}$.
Then the impurity scattering strength is approximately proportional to 
the joint-DOS (JDOS)
\begin{eqnarray}
I^\phi(\q,E)=\sum_{\k,l,m}\rho_{l,m}^\phi(\k,E)\rho_{m,l}^\phi(\k+\q,E)
\label{eqn:QPI},
\end{eqnarray}
where $\rho_{l,m}^\phi(\k,E)=[G_{l,m}^\phi(\k,E+i0_-)-G_{l,m}^\phi(\k,E-i0_+)]/(2i)$
is the QP spectrum.
Authors of previous studies have revealed that the JDOS can simulate the QPI signal\cite{JDOS-QPI-1,JDOS-QPI-2,JDOS-QPI-3}.
Figure \ref{fig:fig5}(e) represents the zero-energy JDOS for $\phi=0$
(without BO) only for the $xy$-orbital FSs, $I^{\phi=0}(\q,0)$.
The JDOS corresponds to the QPI signal by STM measurements. 
Importantly, this simulation result is highly consistent with the autocorrelation map in fig. 2 in Ref. \cite{ARPES-theory}.
The vector $\q_i$ is given by the difference between 
two Fermi points $\k\approx \k_i$ and $\k'\approx -\k_i$.
[Thus, $\q_i\approx 2\k_i$, and $\k_i$ is given in Fig. 1 (c).]
For finite $\phi$, the JDOS becomes anisotropic.
Figure \ref{fig:fig5}(f) shows the
obtained ratios $R_{1-2}=(I_1^\phi-I_2^\phi)/(I_1^\phi+I_2^\phi)$
and $R_{3-4}'=(I_3^\phi-I_4^\phi)/(I_3^\phi+I_4^\phi)$,
where $I_i\equiv I^\phi(\q_i,0)$.
Here, we obtain $I_m^\phi$ ($m=1-4$)
as the maximum value of $I^\phi(\q,0)$ around $\q=\q_i$
because its peak position slightly shifts by $\phi\ne0$.
Because $R_{1-2}$ and $R_{3-4}'$ are negative,
the QPI signal is smaller (larger) at $\q=\q_1,\q_3$ ($\q=\q_2,\q_4$).
(Note that $R_{2-1} =-R_{1-2}$ and $R_{3-4}'=-R_{4-3}'$.)
Here, $R_{1-2}=-0.3 \ (-0.5)$ corresponds to $I_1/I_2=0.54 \ (0.33)$.
Therefore, the QPI signal exhibits sizable nematic anisotropy 
in the $E_{1u}$ BO state for $\phi\sim0.2$eV,
consistent with the experimental results.

In Appendix F, we study the even-parity $E_{2g}$ symmetry BO as the second largest instability of the DW equation. We show that the $E_{2g}$ BO exhibits less anisotropy in the JDOS result and linear relation between the order parameter and FS deformation. Therefore, the odd-parity $E_{1u}$ BO is more consistent with the experiments.

\subsection{VII. NLH EFFECT DUE TO ODD-PARITY BO\vspace{-1em}}
Here, we discuss the NLH effect
as an emergent phenomenon due to the odd-parity BO.
This effect attracts great attention as 
nontrivial nonreciprocal transport in the TRS state
driven by the Berry curvature dipole (BCD)
\cite{NLH-LFu}.
The NLH effect has been recently observed 
in various metals with noncentrosymmetric crystal structures,
such as transition metal dichalcogenides
\cite{NLH-WTe},
twisted bilayer graphene
\cite{NLH-TBG},
and Weyl semimetals
\cite{NLH-review}.
However, the NLH effect driven by odd-parity quantum orders
has rarely been studied so far.
The NLH current due to the BCD is $j_\a=\s_{\a\b\b}E_\b^2$,
and the relation $\s_{xyy}=-\s_{yxy}$ holds.
The NLH conductivity $\s_{\a\b\b}$ is given as \cite{NLH-LFu}
\begin{eqnarray}
\s_{\a\b\b}=\varepsilon_{\a\b z}e^3\tau\frac1N \sum_{b,\k}f(\e_{b,\k}) \d_\b \Omega(b,\k)
\end{eqnarray}
where $\a,\b=x$ or $y$, $f(E)=\{{\rm{exp}}[(E-\mu)/T]+1\}^{-1}$,
and $\tau$ is the conduction electron relaxation time.
here, $\Omega(b,\k)=i[\langle \d_x u_\k^b|\d_y u_\k^b\rangle -(x \leftrightarrow y)]$,
where $\d_\nu \equiv \d/\d k_\nu$ ($\nu=x,y$),
and $u_\k^b$ is the Bloch wavefunction for the $b$th band.
When the inversion symmetry is broken,
the Berry curvature is an odd function of $\k$
[$\Omega(b,\k)=-\Omega(b,-\k)$] in the TRS state
\cite{NLH-LFu}.
Here, we find that the $E_{1u}$ BO induces the 
finite NLH effect due to the BCD.
Importantly, the present NLH effect occurs even without SOC.

\begin{figure}[htb]
\includegraphics[width=.99\linewidth]{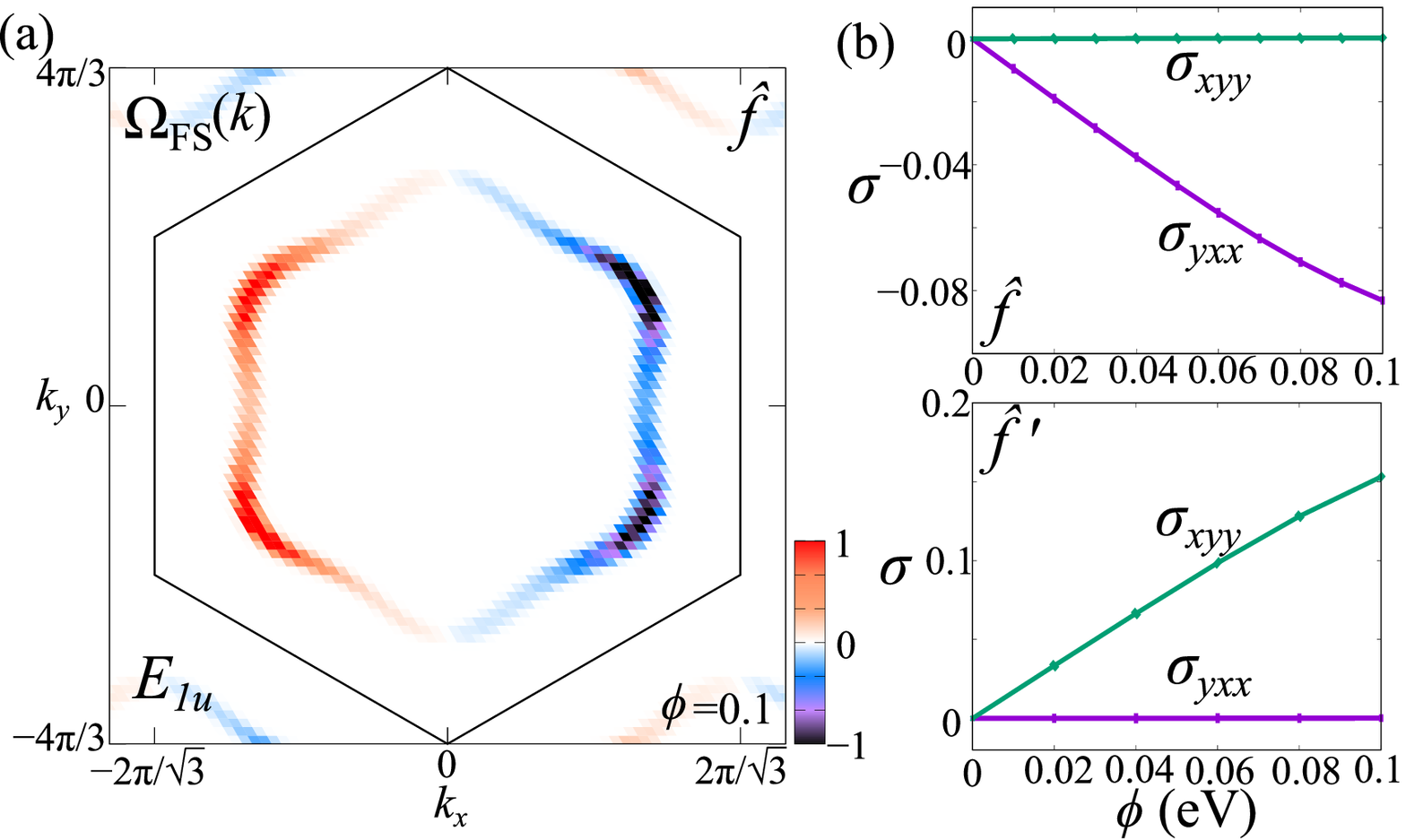}
\caption{
(a) $\Omega_{\rm FS}(\k)$ under $E_{1u}$ bond order (BO) ${\hat f}(\k)$.
(b) Nonlinear Hall (NLH) conductivity induced by the $E_{1u}$ BO 
due to ${\hat f}(\k)$ and ${\hat f}'(\k)$ at $T=0.01$eV.
}
\label{fig:fig6}
\end{figure}

Figure \ref{fig:fig6}(a) shows the Berry curvature on the FS,
$\Omega_{\rm FS}(\k)=\sum_{b}f'(\e_{b,\k})\Omega(b,\k)$
under the $E_{1u}$ BO due to ${\hat f}_{\theta=0}$.
The BCD along the $k_y$ axis originates from 
the mirror symmetry violation with respect to $y \rightarrow -y$ [shown in Fig. \ref{fig:fig4}(b)]. 
The NLH conductivity $\s_{\a\b\b}$ induced by ${\hat f}_{\theta=0}$ BO
is shown in Fig. \ref{fig:fig6}(b).
Here, $\s_{xyy}$ is $\phi$ linear, while $\s_{yxx}=0$.
(Here, we set $\tau=1$ for simplicity.)
We also show the NLH conductivity due to ${\hat f}_{\theta=\pi/2}$ BO.
We stress that the NLH effect in CsTi$_3$Bi$_5$
originates from the quantum phase transition,
in high contrast with the NLH effect in a noncentrosymmetric lattice
\cite{NLH-review,Morimoto-review}.

In addition, the inversion-symmetry violation due to the $E_{1u}$ BO
will be observed by the second-harmonic generation
\cite{SHG-Cd2Re2O7}.
Furthermore, the $E_{1u}$ BO will
give rise to antisymmetric SOC by taking the atomic Ti and Bi SOC,
which will trigger interesting emergent electromagnetism
like the Edelstein effect
\cite{Yanase-JPSJ}.

\subsection{VIII. SUMMARY\vspace{-1em}}
We predict the emergence of the $E_{1u}$-symmetry odd-parity BO,
which has seldom been discussed so far, in CsTi$_3$Bi$_5$.
The emergent phenomena by the $E_{1u}$ BO are very different from 
those by conventional even-parity BO (like the orbital order in FeSe):
The $\q=\0$ $E_{1u}$ BO explains the almost invisible
anomalies in the resistivity and lattice constant at $T\lesssim T_0$
due to $\Delta k_F \propto \phi^2$.
For $T\ll T_0$, however, large $E_{1u}$ BO ($\phi\gg T_0$)
causes the drastic nematic QPI signal observed in CsTi$_3$Bi$_5$
\cite{arXiv:2211.16477,arXiv:2211.12264}.
The result is in good agreement with the spectrum at the Fermi level by ARPES observation in figs. 1(c), 2 and S3 in Ref. \cite{ARPES-theory}.
Furthermore, we reveal that the $E_{1u}$ BO triggers the quantum NLH effect.
In this paper, we have revealed that electric correlations (by paramagnon interference mechanism) cause the odd-parity BO that breaks rotational symmetry in Ti-based kagome metals.


\subsection{ACKNOWLEDGEMENTS\vspace{-1em}}
This paper has been supported by Grants-in-Aid for Scientific
Research from MEXT of Japan (Grants No. JP24K00568, No. JP24K06938, No. JP20K03858, No. JP20K22328, No. JP22K14003, and No. 23H01119), and by the Quantum Liquid Crystal No. JP19H05825 KAKENHI on Innovative Areas from JSPS of Japan.

\subsection{APPENDIX A: BAND CALCULATION FOR CsTi$_3$Bi$_5$\vspace{-1em}}

\begin{figure}[htb]
\includegraphics[width=.99\linewidth]{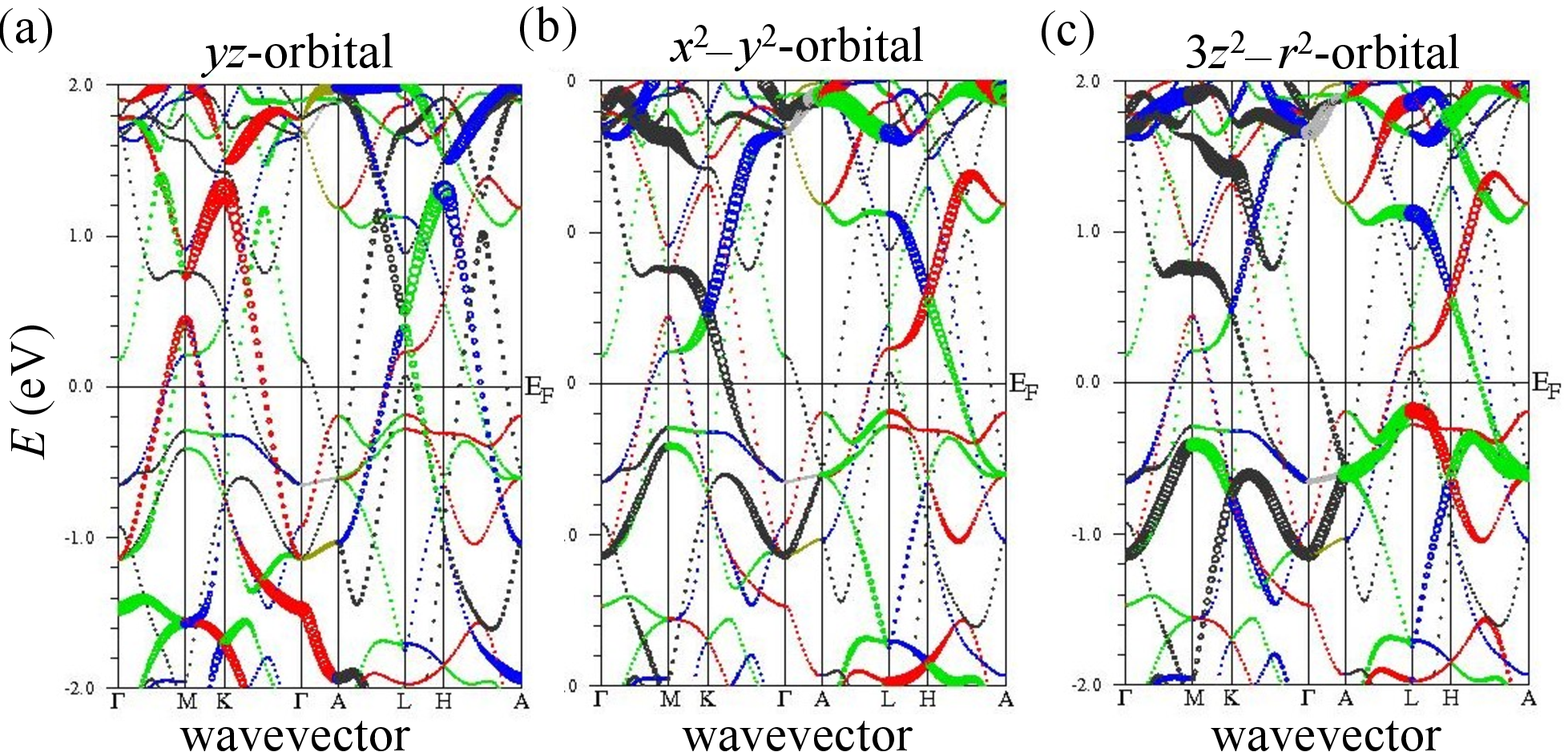}
\caption{
Band structures with (a) $yz$-orbital weight,
(b) $x^2$-$y^2$-orbital weight, and
(c) $3z^2$-$r^2$-orbital weight.
}
\label{fig:fig8}
\end{figure}

Here, we derive the first-principles realistic tight-binding model 
for CsTi$_3$Bi$_5$.
First, we perform the band calculation based on the {\footnotesize WIEN}2{\footnotesize K} software,
by using the crystal structure reported in Ref. \cite{arXiv:2209.11656}.
The FSs are essentially 2D 
because the interlayer hopping integrals are small.
The large cylindrical FS around the $\Gamma$ point 
is mainly composed of the $xy$ orbital of Ti $3d$ electrons.
The two cylindrical FSs around $K$, $K'$ points
are composed of $xz$-orbital electrons.
Thus, the major FSs of CsTi$_3$Bi$_5$ are mainly composed of
two $d$ orbitals ($xy$ and $xz$) of Ti ions on sublattices A, B, and C.
The band structures with $xy$- and $xz$-orbital weights are shown in 
Figs. \ref{fig:fig7}(a) and \ref{fig:fig7}(b), respectively.
Each $d$-orbital DOS is shown in Fig. \ref{fig:fig7}(c).
Other less important $d$-orbital weights
are shown in Figs. \ref{fig:fig8}(a)$-$\ref{fig:fig8}(c).

In the obtained band structure, the SOC is neglected.
The effect of SOC is large for the Bi $6p$-orbital band
that gives the small Fermi pocket around the $\Gamma$ point,
while the major $3d$-orbital bands are affected by SOC 
only slightly \cite{arXiv:2209.11656}.
Therefore, the effect of SOC is safely neglected.
We just introduce the shift of the $xy$-orbital level 
$\delta E_{xy}=-0.15$eV to represent the self-hole-doping
due to SOC-induced enlargement of the $6p$-orbital 
electron pocket around the $\Gamma$ point.

Next, we derive the 30 orbital tight-binding model 
by using {\footnotesize Wannier}90 software.
In the main text, we use the 2D model 
by neglecting the small interlayer hopping integrals.
Then we perform the RPA using the 30 orbital model. To find out the significant $d$ orbitals with strong electron correlations, we first applied the Coulomb interaction $H_U$ only on the $d_{xz}$ and $d_{xy}$ orbitals (as the multiorbital Coulomb interaction mentioned in the main text).
We revealed that the spin Stoner factor $\a_S$ reaches unity 
when the intraorbital term $U=1.62$eV (RPA), 
where the interorbital term $U'=U-2J$
and the exchange term $J=0.1U$ are used 
for the six-orbital (3$b_{3g}$+3$b_{2g}$) Coulomb interaction.
When $\a_S\gtrsim0.9$, the spin susceptibility develops
only for the $xy$ orbital $\chi^s_{xy}(\q)$.

It is verified that both $\a_S$ and $\chi^s_{xy}(\q)$
are well reproduced even if only the intra-$xy$-orbital 
Coulomb interaction $U_{xy}=2.18$eV (RPA) is considered 
in the numerical study(as the single-orbital Coulomb interaction mentioned in the main text).
[If only the intra-$xz$-orbital interaction $U_{xz}$ is considered,
large $U=5.29$eV (RPA) is needed for realizing $\a_S=1$.]
Therefore, the spin fluctuations in a Ti-based kagome metal are 
highly orbital selective.
In the main text, we study nematic order due to the 
paramagnon interference mechanism.
Since the paramagnon develops only on the $xy$ orbital,
we introduce only the $xy$-orbital Coulomb interaction
in the FLEX and DW equation analyses in the main text.

\subsection{APPENDIX B: FLEX SELF-ENERGY and DW EQUATION\vspace{-1em}}

\begin{figure}[htb]
\includegraphics[width=.99\linewidth]{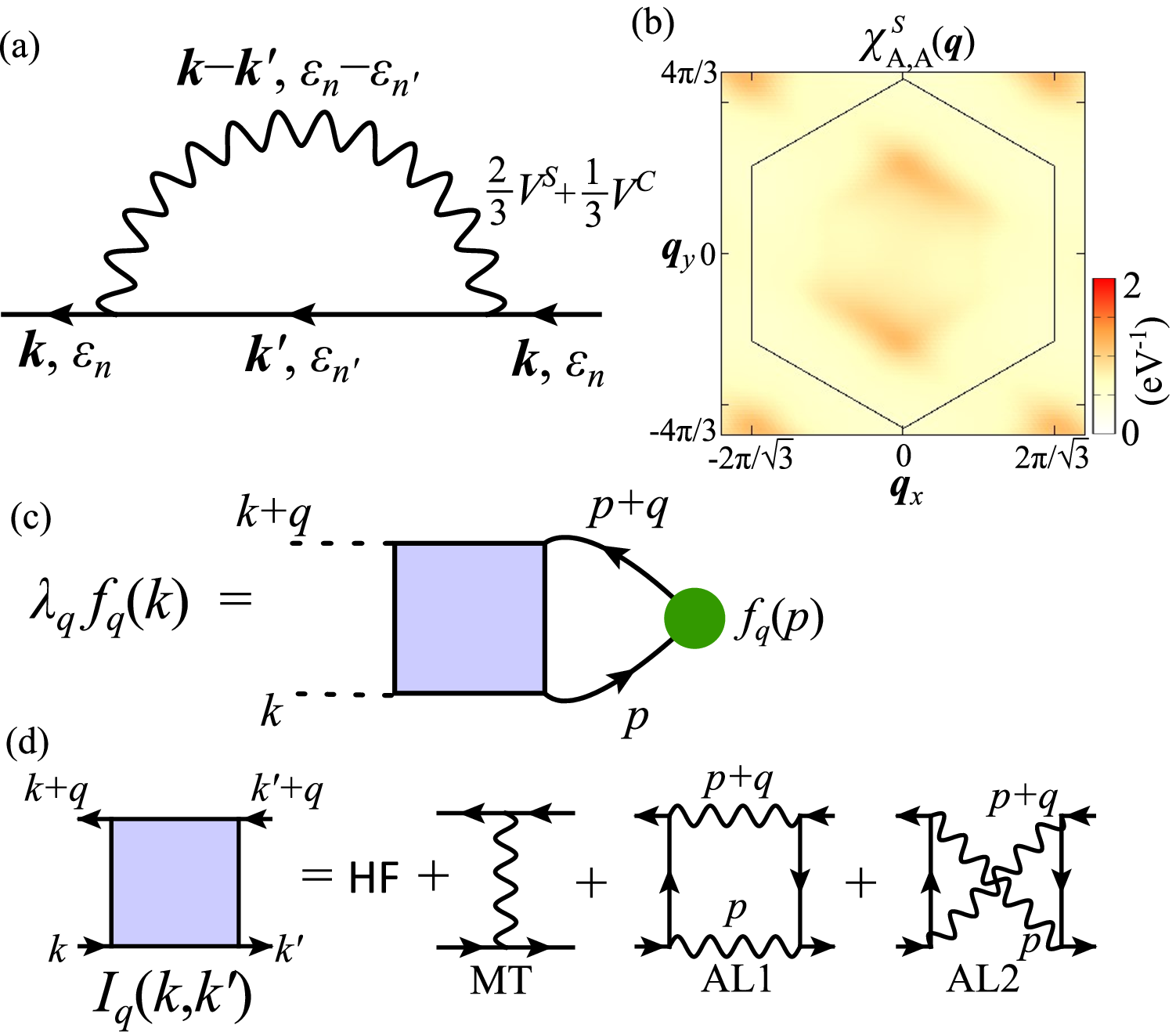}
\caption{
(a) Spin susceptibility $\chi^s_{A,A}(\q)$ by the fluctuation-exchange (FLEX) calculation for the multiorbital Coulomb interaction.
(b) Diagrammatic expression of the self-energy.
(c) Diagrammatic expression of the density-wave (DW) equation.
(d) Kernel function $I$ composed of one Hartree term,
one Maki-Thompson (MT) term, and two Aslamazov-Larkin (AL) terms.
Note that simple local charge density order is prohibited by the Hartree term.
}
\label{fig:fig9}
\end{figure}

In the main text,
we calculate the spin susceptibility 
by including the self-energy effect self-consistently.
For this purpose, we use the FLEX approximation
\cite{Bickers2,Kontani-RH,Kontani-rev1}
\begin{eqnarray}
\Sigma_{l,m}(k)&=&\frac{T}{N}\sum_{q}G_{l,m}(k-q)V_{l,m}(q),
\label{eqn:SigmaS} \\
V_{l,m}(q)&=&\frac{U^2}{2}[3\chi^s_{l,m}(q)+\chi^c_{l,m}(q)-\chi^0_{l,m}(q)],
\label{eqn:VS} \\
\chi^0_{l,m}(q)&=&-\frac{T}{N}\sum_k G_{l,m}(k+q)G_{m,l}(k),
\label{eqn:chi0S} \\
{\hat\chi}^{s(c)}(q)&=&{\hat \chi}^0(q)[{\hat 1}-(+)U{\hat \chi}^0(q)]^{-1}
\label{eqn:chiS} 
\end{eqnarray}
where 
the indices $l,m$ represent the $xy$ orbital at sublattices A, B, and C;
$k\equiv[\k,\e_n=(2n+1)\pi T]$; and 
$q\equiv(\q,\w_l=2l\pi T)$.
Here, $\Sigma_{l,m}(k)$ is the self-energy
shown in Fig. \ref{fig:fig9}(a),
$\chi^{s(c)}_{l,m}(q)$ is the spin (charge) susceptibility,
and $G_{l,m}(k)$ is the Green's function on the $xy$ orbital.
Also, $U$ is the Coulomb interaction on the $xy$ orbital.
Here, 
$G_{l,m}(k)=[(i\e_n+\mu){\hat 1}-{\hat h}^0_\k-{\hat \Sigma}(k)]_{l,m}^{-1}$,
where ${\hat h}^0_\k$ is a $30 \times 30$ matrix expression of the 
kinetic term given by the Fourier transform of the 
present tight-binding model.
[Note that the matrix elements of ${\hat \Sigma}(k)$ 
are zero except for three $xy$ orbitals.]
Here, we solve Eqs. (\ref{eqn:SigmaS})$-$(\ref{eqn:chiS})
self-consistently.
In the numerical study, we use $60\times60$ $\k$-meshes
and $8192$ Matsubara frequencies. We verified the $\k$-mesh up to $120\times120$, and the change of $\alpha_S$ and $\lambda$ is $\sim0.1\%$. Therefore, the numerical calculation by the $60\times60$ $\k$-mesh is well converged in this FLEX study.

Additionally, we performed the FLEX calculation for the multiorbital case and exhibit it in Fig. \ref{fig:fig9}(b). The change of the spin fluctuations between the single- and multiorbital calculations is small. Therefore, it is sufficient to study the single-orbital FLEX in this paper.

Figure \ref{fig:fig9}(c) shows the
diagrammatic expression of the DW equation, 
which is given in Eq. (\ref{eqn:DWeq}).
The kernel function $I$, given by the Ward identity 
($I=\delta\Sigma/\delta G$), is composed of
one single-magnon exchange term and 
two double-magnon interference terms.
The former is called the MT term,
and the latter are called the AL terms.
Its derivation based on the LW free-energy theory 
is given in Ref. \cite{Tazai-LW},
and the expression of it is in Appendix D.
We verified that the AL (MT) contribution
to the $E_{1u}$ state is $\lambda_{E_{1u}}^{\rm AL}=0.75$ 
($\lambda_{E_{1u}}^{\rm MT}=0.27$).
It is noteworthy that $\lambda_{E_{1u}}^{\rm Hartree}=0$.

Here, we explain the essential role of the AL terms
for the charge channel DW,;
${\hat f}^c \equiv ({\hat f}^\uparrow+{\hat f}^\downarrow)/2$.
The kernel functions for AL1 ($I^+$) and AL2 ($I^-$)
are approximately given as
$I_\q^\pm(k,k')\approx T\sum_p \frac{3U^4}{2}G(k\pm p)G(k'-p)X_\q(p)$,
where $X_\q(p)=\chi^s(p)\chi^s(p+\q)$.
The present interference between two paramagnons
gives rise to the charge-channel DW constructively
in the DW equation.
Importantly, this process does not contribute to the spin-channel DW,
${\hat f}^s \equiv ({\hat f}^\uparrow-{\hat f}^\downarrow)/2$,
because the two-paramagnon process preserves TRS
\cite{Kontani-AdvPhys}.
In FeSe, the AL terms drive nonmagnetic nematic order.
The importance of AL terms was
verified by the fRG study
\cite{Kontani-AdvPhys}.

\begin{figure}[htb]
\includegraphics[width=.99\linewidth]{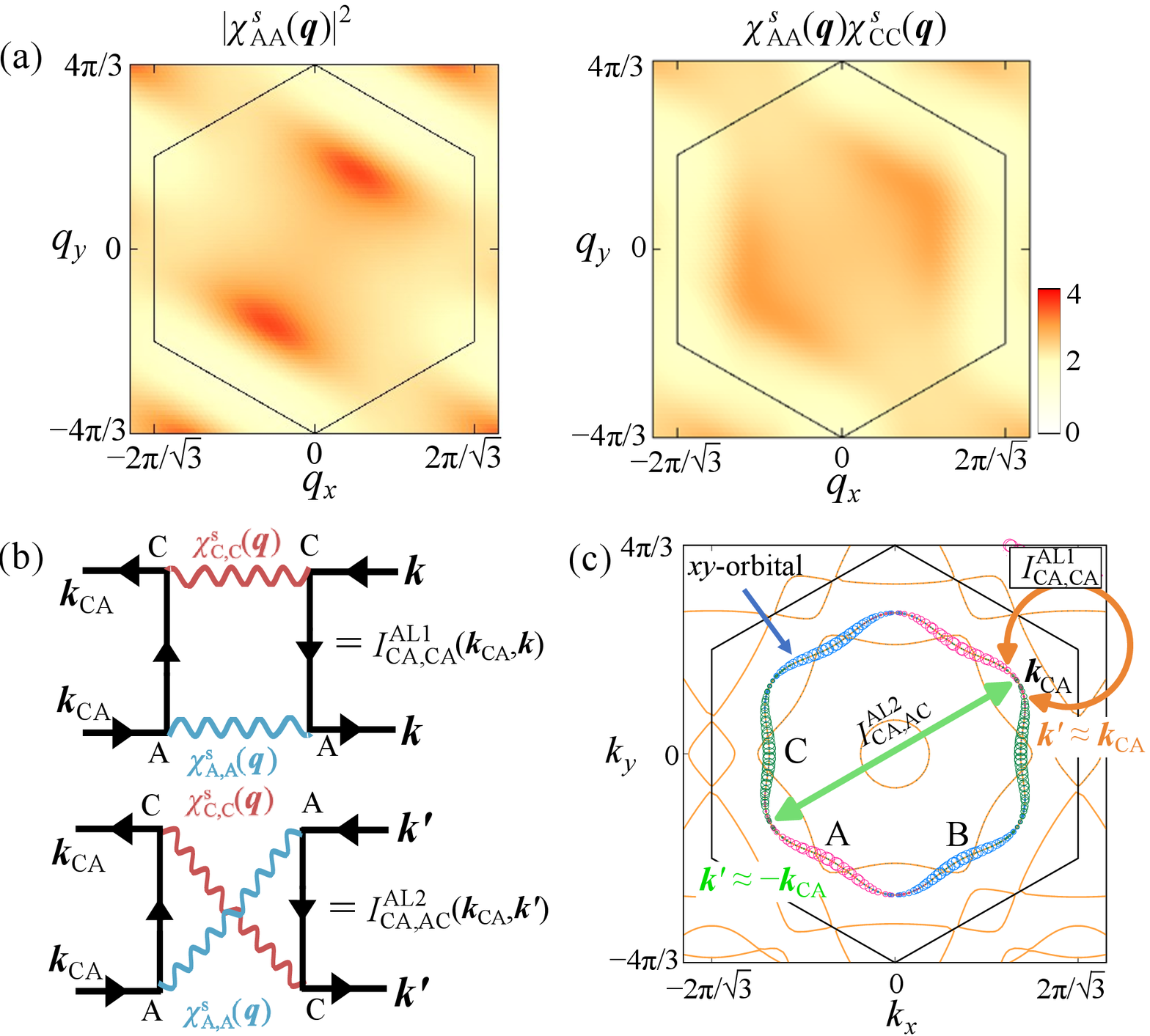}
\caption{
(a) $X_{\rm A,A}(\q)\equiv\{\chi^s_{\rm A,A}(\q)\}^2$ and 
$X_{\rm A,C}(\q)\equiv \chi^s_{\rm A,A}(\q)\chi^s_{\rm C,C}(\q)$.
(b) AL1 term $I^{\rm AL1}_{\rm CA,CA}(\k_{\rm CA},\k)$ and 
AL2 term $I^{\rm AL2}_{\rm CA,AC}(\k_{\rm CA},\k')$.
(c) Attraction given by AL1 and AL2 terms.
The weights of the sublattices A (red), B (blue), and C (green) are shown.
}
\label{fig:figS5-2}
\end{figure}

Below, we discuss why the AL terms give the $E_{1u}$ BO 
in a Ti-based kagome metal with the sublattice degrees of freedom.
The AL terms in Fig. \ref{fig:fig9}(d) 
for the form factor $f^{l,m}$ are approximately proportional to
the convolution  
$X_{lm}(\q)=\frac1N \sum_\p\chi^s_{l,l}(\p)\chi^s_{m,m}(\p+\q)$
($l,m={\rm A,B,C}$).
In general, the eigenvalue at $\q$ is roughly proportional to $X_{lm}(\q)$,
which is large when $\q\sim{\bm0}$,
and the peak (the width) of $\chi^s(\p)$ is high (broad).
Figure \ref{fig:figS5-2}(a) shows
$\{\chi^s_{\rm A,A}(\q)\}^2$ and 
$\chi^s_{\rm A,A}(\q)\chi^s_{\rm C,C}(\q)$ in the present model at $\a_S=0.9$.
Due to the broadness of the peak of $\chi^s_{\rm A,A}(\q)$ 
shown in Fig. \ref{fig:fig1}(d),
$\frac1N \sum_\q \chi^s_{\rm A,A}(\q)\chi^s_{\rm C,C}(\q)=1.81$
is as large as 
$\frac1N \sum_\q \{\chi^s_{\rm A,A}(\q)\}^2=1.83$.
This situation is favorable for the $E_{1u}$ BO
given by the intersublattice form factor
shown in Fig. \ref{fig:fig4}(a).


The symmetry of the form factor is determined by the 
momentum dependence of $I(\k,\k')$.
Figure \ref{fig:figS5-2}(b) shows 
the AL1 term $I^{\rm AL1}_{\rm CA,CA}(\k_{\rm CA},\k)$ and 
AL2 term $I^{\rm AL2}_{\rm CA,AC}(\k_{\rm CA},\k')$.
As we explained in Refs. \cite{Kontani-sLC,Onari-b2g},
the AL1 term gives the attractive interaction for $\k\approx\k_{\rm CA}$.
In contrast, the AL2 term gives the attractive interaction 
for $\k'\approx-\k_{\rm CA}$,
as depicted in Fig. \ref{fig:figS5-2}(c).
Such $\k$ dependence originates from the 
p-h pair $G(p)G(p)$ in the AL1 term
and the particle-particle pair $G(p)G(-p)$ in the AL2 term
\cite{Kontani-sLC,Onari-b2g}.
Both [AL1] and [AL2] cooperatively contribute to the 
odd-parity BO, as we explain in the main text.

\subsection{APPENDIX C: FORM FACTOR AND SYMMETRY BREAKING IN SELF-ENERGY\vspace{-1em}}

In the strongly correlated metals,
various kinds of the DW orders emerge,
such as spin/charge density waves,
even/odd-parity BOs, and charge/spin current orders.
These DW orders are represented as symmetry breaking 
in the self-energy \cite{Tazai-LW}.
Here, we focus on the DW at $\q={\bm0}$.
By following Ref. \cite{Tazai-LW}, we divide the self-energy into 
\begin{eqnarray}
{\hat \Sigma}(k)={\hat \Sigma}^0(k)+\delta {\hat t}(k) ,
\label{eqn:self-devide}
\end{eqnarray}
where ${\hat \Sigma}^0$ is the normal-state self-energy
without any symmetry breaking,
and $\delta {\hat t}$ is equal to the DW order parameter 
introduced in Appendix D.
Here, $\Sigma^0$ belongs to $A_{1g}$ symmetry, while 
$\delta t$ belongs to non-$A_{1g}$ symmetry.
Thus, $\delta t=0$ for $T>T_{\rm c}$.
Hereafter, we denote $\delta {\hat t}(k) \equiv \phi {\hat f}(k)$,
where ${\hat f}(k)$ is the form factor that is normalized as
$\max_{l,m,k} |f_{lm}(k)|=1$.
The form factor is microscopically obtained by solving the DW equation.
Another equivalent interpretation of the form factor $f_\q^L(k)$ is 
the p-h condensation 
$\sum_\s \{ \langle c_{\k+\q,l,\s}^\dagger c_{\k,l',\s}\rangle - \langle \cdots \rangle_0 \}$
\cite{Tazai-LW}.


Because we consider the thermal equilibrium state,
the form factor satisfies the Hermitian condition
$f_\q^{l,l'}(k)=[f_{-\q}^{l',l}(k+\q)]^*$.
Here, $l$ and $l'$ are the orbital and the sublattice,
and $\q$ is the wavevector of the DW.
This condition is directly derived from the Hermitian condition
for the hopping integral between sites $i$ and $j$:
$\delta t_{i,j}^{l,l'}= (\delta t_{j,i}^{l,l'})^*$.
In the BO without TRS breaking,
$\delta t_{i,j}^{l,l'}=\delta t_{j,i}^{l',l}=$real.
In the case of $E_{1u}$ BO,
the form factor $({\hat f},{\hat f}')$ satisfies the relation
$f_{\rm AC}(-\k_{\rm CA})=-\{f_{\rm CA}(\k_{\rm CA})\}^*=f_{\rm CA}(\k_{\rm CA})$.
For any form factor  ${\hat f}$,
$x{\hat f}$ violates the Hermitian condition if $x$ is not real.
Therefore, the linear combination of the form factor is always given as
${\hat f}_\theta= {\hat f} \cos\theta + {\hat f}'\sin\theta$.
This fact is very different from the superconducting state 
in a 2D irreducible representation (irrep; $\Delta,\Delta'$), where the chiral state 
($\Delta+i\Delta'$) without TRS is allowed.

Meanwhile, the DW state without TRS occurs when
$\delta t_{i,j}^{l,l'}=-\delta t_{j,i}^{l',l}$ is imaginary.
This charge current order is actively discussed
in V-based kagome metals \cite{Tazai-kagome2}.


\subsection{APPENDIX D: DW EQUATION FORMULA\vspace{-1em}}

Here, we rewrite $\delta t_{k\sigma}$ as
\begin{eqnarray}
\delta t_{k\sigma} \equiv  \phi f_{k\sigma} ,
\end{eqnarray}
where $\phi$ is a real parameter, and $f_{k\sigma}$ is 
the normalized order parameter that belongs to one of the 
irreps in non-$A_{1g}$ symmetry.
It is convenient to set $\max_k |f_{k\sigma}|=1$ 
because the relation $\phi=\max_k|\delta t_{k\sigma}|$ holds.

The order parameter $f_{k}^{q}$ is derived from the DW equation:

\begin{eqnarray}
\lambda f_{k\sigma}= -\frac{T}{N} \sum_{k'\sigma' }  I^{\sigma \sigma'}_{kk'}
(G^{0}_{k'\sigma'})^2  f_{k'\sigma'} ,
\label{eqn:cons06q0}
\end{eqnarray} 
where we denote the kernel function
$I^{\sigma \sigma'}_{kk'} \equiv \left. I^{\sigma \sigma'}_{kk'} \right|_{\Sigma^0}$
to simplify the notation. From now on, we omit the spin notation for simplification. When $\q=0$, $I^{L,M}_{k,k';\q}$ is given by the Ward identity 
$I^{L,M}_{\q}=-\delta \Sigma^L(k)/\delta G^M(k')$, where $L\equiv (l,l')$, $M\equiv(m,m')$ represent the pair of sublattice indices A, B, and C. 
It is the irreducible four-point vertex consisting of one Hartree term, one single-magnon exchange MT term, and two double-magnon interference AL1 and AL2 terms. The analytic expression is given as

\begin{align}
&I^{l,l',m,m'}_{\q}(k,k')=\Gamma^c_{l,l',m,m'} \nonumber\\
&-\sum_{b=s,c}\frac{a^b}2 [V^b_{l,m;l',m'}(k-k')\nonumber\\
&-T\sum_p\sum_{l_1,l_2,m_1,m_2}V^b_{l,l_1;m,m_2}(p_+)V^b_{m',m_2;l',l_2}(p_-)\nonumber\\
&\times G_{l_1,l_2}(k-p)G_{m_2,m_1}(k'-p)\nonumber\\
&-T\sum_p\sum_{l_1,l_2,m_1,m_2}V^b_{l,l_1;m_2,m'}(p_+)V^b_{m_1,m;l',l_2}(p_-)\nonumber\\
&\times G_{l_1,l_2}(k-p)G_{m_2,m_1}(k'+p)]
\label{eqn:kernel function} ,
\end{align}
where the double counting in the first- and second-order terms should be subtracted.
Here, $a^{s(c)}=3(1),p_{\pm}\equiv p+\q/2,p=(p,\omega_l)$, 
and $\hat V^b$ is the $b$-channel interaction matrix given by $\hat V^b=\hat\Gamma^b+\hat\Gamma^b\hat\chi^b\hat\Gamma^b$. 
Also, $\hat\Gamma^b$ is the $b$-channel bare multiorbital Coulomb interaction. 
[In the present model, $(\hat\Gamma)^s_{ll'mm'}=U\delta_{ll'}\delta_{l'm}\delta_{mm'}$ and $(\hat\Gamma)^c_{ll'mm'}=-U\delta_{ll'}\delta_{l'm}\delta_{mm'}$.]The susceptibility ${\hat \chi}^b(q)$ is given in the main text.
The first term in Eq. (\ref{eqn:kernel function}) is the MT term when the second and third terms are the AL terms.

Here, we should stress the importance of the AL terms. 
In previous studies, it has been proved that, although the fRG method includes higher-order terms, it predicted the same BO \cite{Tsuchiizu4,fRG-BEDT} or orbital order \cite{Tsuchiizu1} in both single-orbital \cite{fRG-sin-orb,fRG-BEDT} and two-orbital Hubbard models \cite{Tsuchiizu1,Tsuchiizu4} as the DW equation did.
Therefore, the validity of the DW equation and the essence of the AL terms were verified.

In Eq. (\ref{eqn:cons06q0}),
the largest eigenvalue $\lambda$ reaches unity at $T=T_{\rm c}$,
and its eigenvector gives the form factor of the DW state.
In Ref. \cite{Tazai-LW},
the authors discussed the Ginzburg-Landau (GL) free energy 
based on the Luttinger-Ward-Potthoff theory.
The $\q$-dependent second-order GL coefficient is simply given by the
form factor and the eigenvalue derived from 
the DW Eq. (\ref{eqn:cons06q0}).




\subsection{APPENDIX E: NEMATIC FS DEFORMATION BY $E_{1u}$ BO FOR $\theta\ne0$\vspace{-1em}}

\begin{figure}[htb]
\includegraphics[width=.99\linewidth]{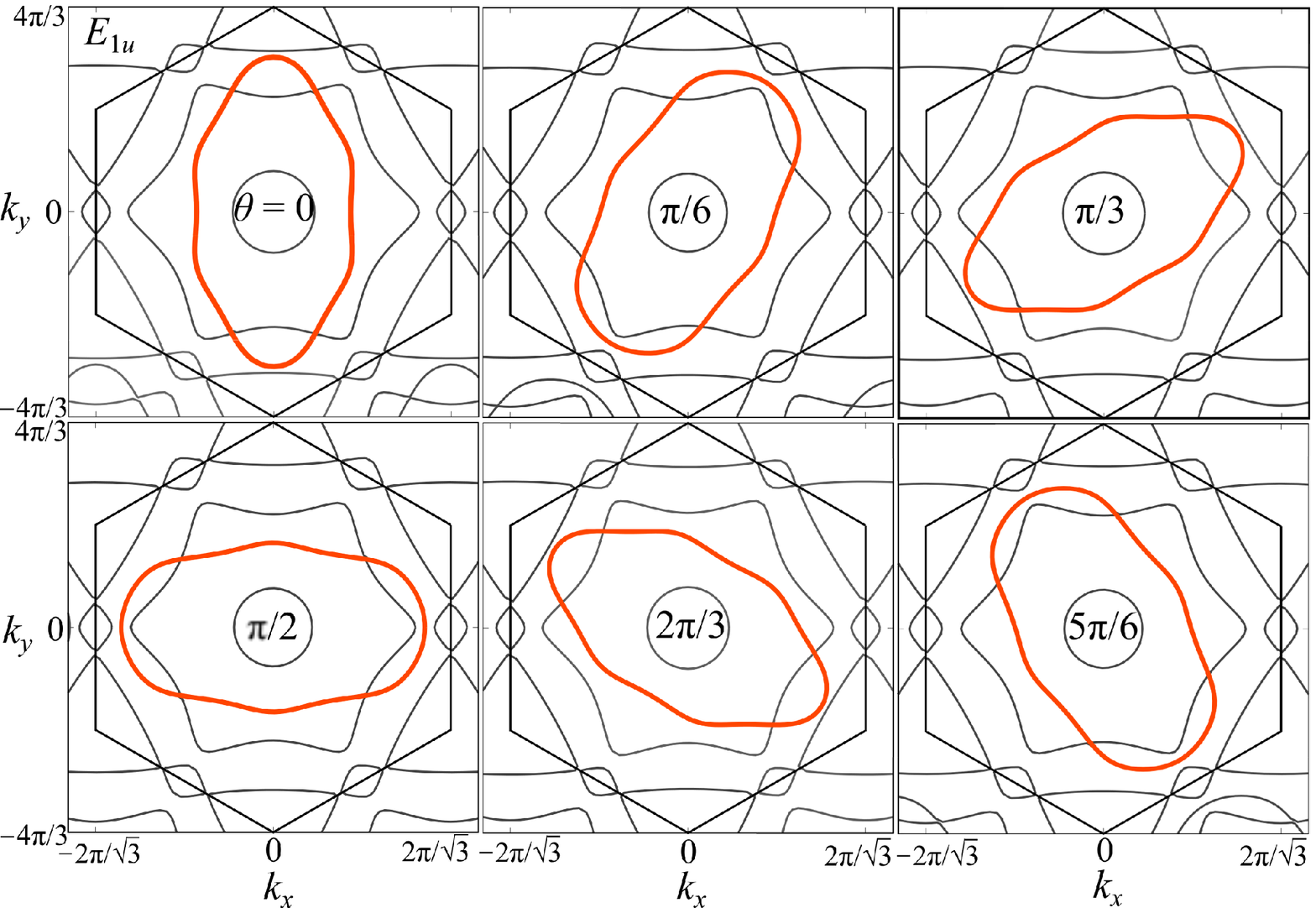}
\caption{
Nematic Fermi surfaces (FS) at $\theta=0,\pi/6,\dots,5\pi/6$
in the $E_{1u}$ bond-order (BO) state at $\phi=1$.
}
\label{fig:fig11}
\end{figure}

We discuss that the present $E_{1u}$ form factor
${\hat f}_\theta\equiv {\hat f}\cos\theta + {\hat f}' \sin \theta$
gives the nematic FS deformation.
Here, we introduce the symmetry breaking in the self-energy as
$\delta{\hat t}_\k^\theta=\phi {\hat f}_\theta(\k)$.
The FSs are derived from the eigenvalues of 
${\hat h}_\k^0+\delta{\hat t}_\k^{\theta}$.
Because ${\hat f}_\theta$ belongs to $E_{1u}$ representation,
the FS deformation is very small since it is proportional to $\phi^2$.
For this reason, here, we set $\phi\sim1$ to exaggerate the deformation.
Figure \ref{fig:fig11} is the obtained FSs
for $\theta=0,\pi/6,\pi/3,\pi/2,2\pi/3$, and $5\pi/6$.
We stress that the FS for $\theta$ is equal to the FS for $\theta+\pi$
because the FS deformation due to the $E_{1u}$ BO 
is proportional to $\phi^2$.
For this reason, the induced lattice deformation and 
kink in the resistivity at $T_0$ would be quite small, 
consistent with experimental reports.

\subsection{APPENDIX F: EVEN-PARITY $E_{2g}$ BO STATE AT $\q={\bm0}$\vspace{-1em}}

\begin{figure}[htb]
\includegraphics[width=.8\linewidth]{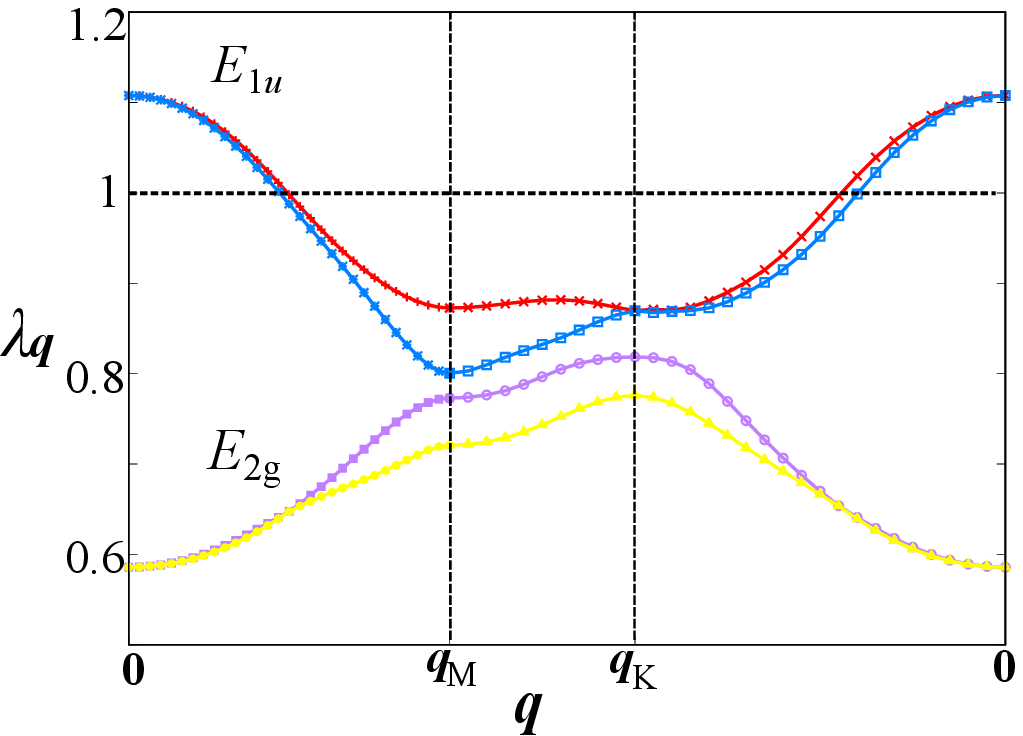}
\caption{
First to fourth eigenvalues for $U=4$eV as functions of $\q$.
The first and second largest eigenvalues correspond to the $E_{1u}$ bond-orders (BOs),
and the third and fourth correspond to the $E_{2g}$ BOs. 
}
\label{fig:four-eigenvalues}
\end{figure}

In the present DW equation analysis, 
the $\q={\bm0}$ $E_{1u}$ BO solution is obtained as the largest eigenvalue.
Figure \ref{fig:four-eigenvalues} shows the
first to fourth eigenvalues for $U=4$eV as functions of $\q$.
The first and second largest eigenvalues correspond to the $E_{1u}$ BOs,
and the third and fourth ones correspond to the $E_{2g}$ BOs. 
The obtained eigenvalue at $\q={\bm0}$ is $1.11$ ($0.59$)
for the $E_{1u}$ ($E_{2g}$) BO solution at $U=4$eV and $T=0.01$eV.
Thus, the obtained $\lambda_{E_{2g}}$ is $\sim0.5$ 
smaller than $\lambda_{E_{1u}}$ because its form factor
$g^{\rm C,A}(\k)\propto \cos \k\cdot{\bm a}_{\rm CA}$
is smaller than the $E_{1u}$ form factor $f^{\rm C,A}(\k)$
in magnitude at $\k\sim\k_{\rm CA}$; see Fig. \ref{fig:fig4} (a).
However, both $E_{1u}$ and $E_{2g}$ BOs may appear
by modifying the model parameters.

\begin{figure}[htb]
\includegraphics[width=.99\linewidth]{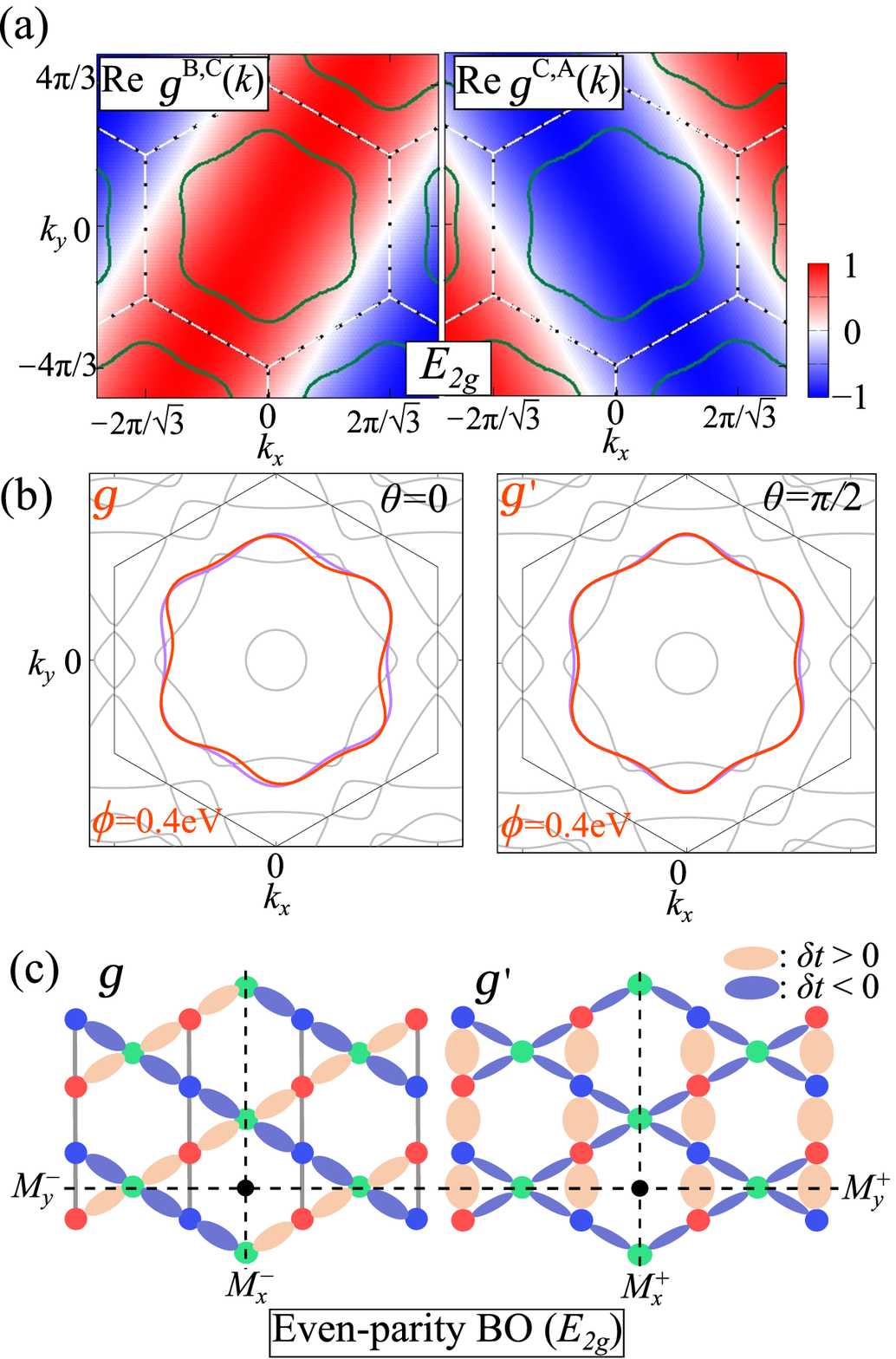}
\caption{
(a) Form factor of even-parity $E_{2g}$ bond-order (BO), 
which corresponds to the second largest eigenvalue at $\q={\bm0}$.
(b) Nematic Fermi surfaces (FSs) in the $E_{2g}$ BO (${\hat g},{\hat g}'$)
at $\theta=0$ and $\pi/2$ for $\phi=0.4$eV.
Even-parity $E_{2g}$ BO form factor 
(c) ${\hat g}$ and (d) ${\hat g}'$.
They are apparently different from the odd-parity $E_{1u}$ BOs 
${\hat f}$ and ${\hat f}'$ in Fig. \ref{fig:fig4}.
}
\label{fig:fig13}
\end{figure}

The $E_{2g}$ BO solution is doubly degenerate,
so its form factor is given by the linear combination of 
two orthogonal functions ${\hat g}_\k$ and ${\hat g}'_\k$;
that is,
${\hat g}_\theta\equiv {\hat g}\cos\theta + {\hat g}'\sin \theta$.
Figure \ref{fig:fig13}(a) shows the obtained ${\hat g}_\k$ 
for the $E_{2g}$ BO.
The realized nematic FS deformation is shown in 
Fig. \ref{fig:fig13}(b).
In the $E_{2g}$ BO,
the director of the nematic FS is parallel to 
$\{\cos[(\theta+\pi/2)/2],\sin[(\theta+\pi/2)/2]\}$.
(Note that the director is parallel to $(\cos\theta,\sin\theta)$
in the $E_{1u}$ BO.)

The $E_{2g}$ BO form factor in real space is shown in 
Figs. \ref{fig:fig13}(c) and \ref{fig:fig13}(d).
Each $E_{2g}$ BO form factor is invariant under the inversion operation.
Thus, the fluctuations of the ferro-$E_{2g}$ BO
can be measured as the development of nematic susceptibility.
In contrast, the ferro-$E_{1u}$ BO fluctuations cannot be observed by the nematic susceptibility measurement because it does not form bilinear coupling with any share modulus.
In fact, the $E_{1u}$ BO form factors shown in Figs. \ref{fig:fig4}(b) and \ref{fig:fig4}(c) 
change their sign under the inversion operation,
called the electric toroidal quadrupole order or electric octupole order.




\subsection{APPENDIX G: $\Delta k_F$, $\Delta v_F$, AND ANISOTROPIC QPI SIGNAL DUE TO EVEN-PARITY BO\vspace{-1em}}

\begin{figure}[htb]
\includegraphics[width=.99\linewidth]{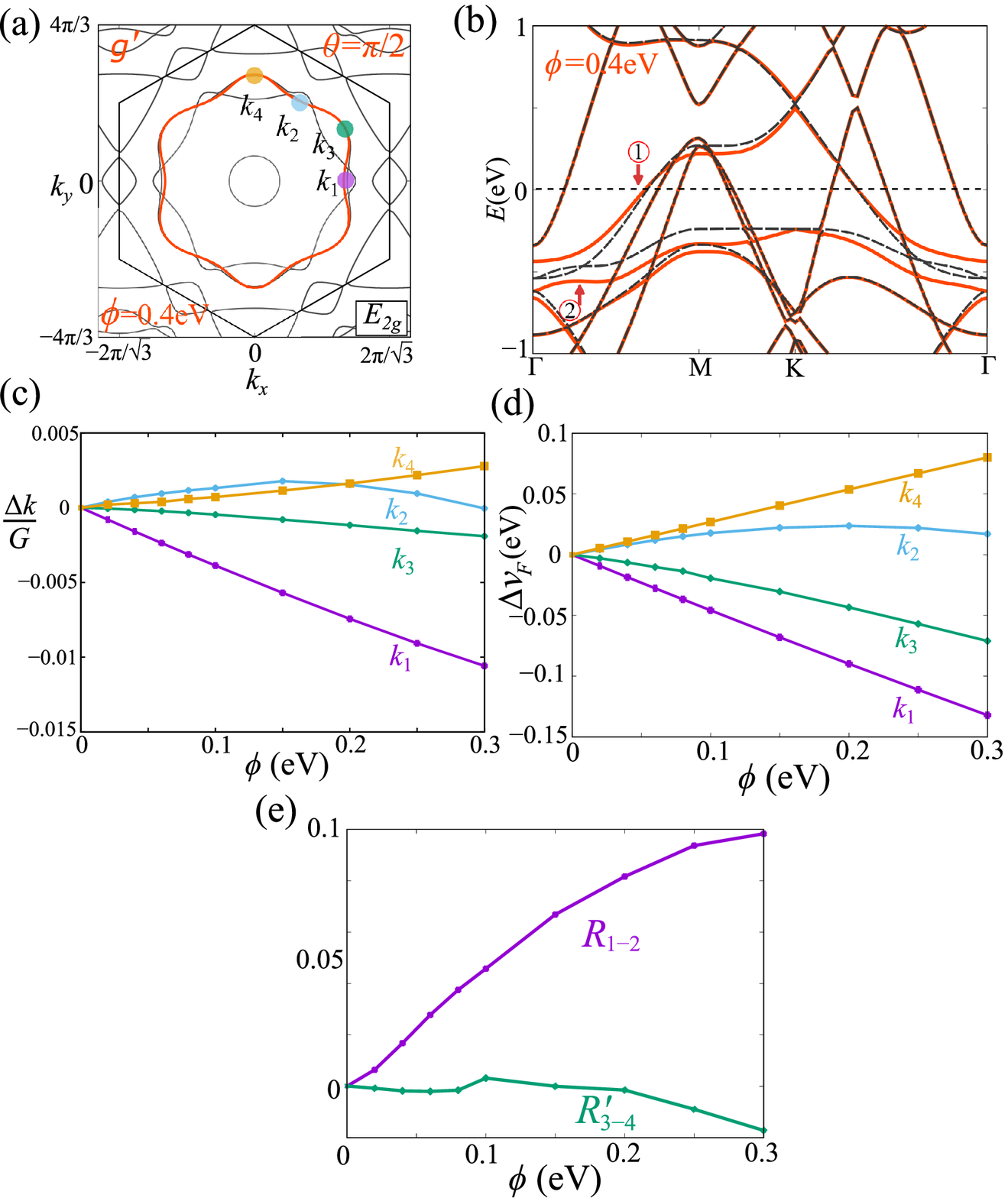}
\caption{
(a) Fermi surface (FS) deformation due to the $E_{2g}$-symmetry self-energy
$\delta{\hat t}_\k^\theta=\phi {\hat g}_\theta(\k)$
for $\phi=0.4$eV at $\theta=\pi/2$.
(b) Band dispersion deformation at $\phi=0.4$eV.
(c) Deformation of the Fermi momentum $\Delta k_F/G$,
where $G=4\pi/\sqrt{3}$.
(d) Deformation of the Fermi velocity $\Delta v_F$.
(e) Anisotropy of the quasiparticle interference (QPI) intensity 
$R_{1-2}\equiv(I_1^\phi-I_2^\phi)/(I_1^\phi+I_2^\phi)$
and $R_{3-4}'\equiv(I_3^\phi-I_4^\phi)/(I_3^\phi+I_4^\phi)$ as a function of $\phi$.
The obtained nematic anisotropy by $E_{2g}$ bond order (BO) is smaller than experimental result.
}
\label{fig:fig-def-even}
\end{figure}

Now we analyze the $E_{2g}$ BO state.
The deformations of the FS and the band dispersion
$\Delta k_F/G$, $\Delta v_F$ 
are shown in Figs. \ref{fig:fig-def-even}(a)$-$\ref{fig:fig-def-even}(d), respectively.
They are proportional to $\phi$ in the $E_{2g}$ BO.
Figure \ref{fig:fig-def-even}(e) is the
anisotropy of the QPI signal due to the $E_{2g}$ BO state
$R_{1-2}=(I_1^\phi-I_2^\phi)/(I_1^\phi+I_2^\phi)$
and $R_{3-4}'=(I_3^\phi-I_4^\phi)/(I_3^\phi+I_4^\phi)$,
where $I_i\equiv I^\phi(\q_i,0)$, and the definition of $\q_i$ is the same as in the main text.
Here, $R_{1-2}=0.1$ corresponds to $I_1/I_2=0.82$.
The obtained nematic anisotropy by the $E_{2g}$ BO is 
smaller than that by the $E_{1u}$ BO shown in Fig. \ref{fig:fig-def-even}(f).

\begin{figure}[htb]
\includegraphics[width=.99\linewidth]{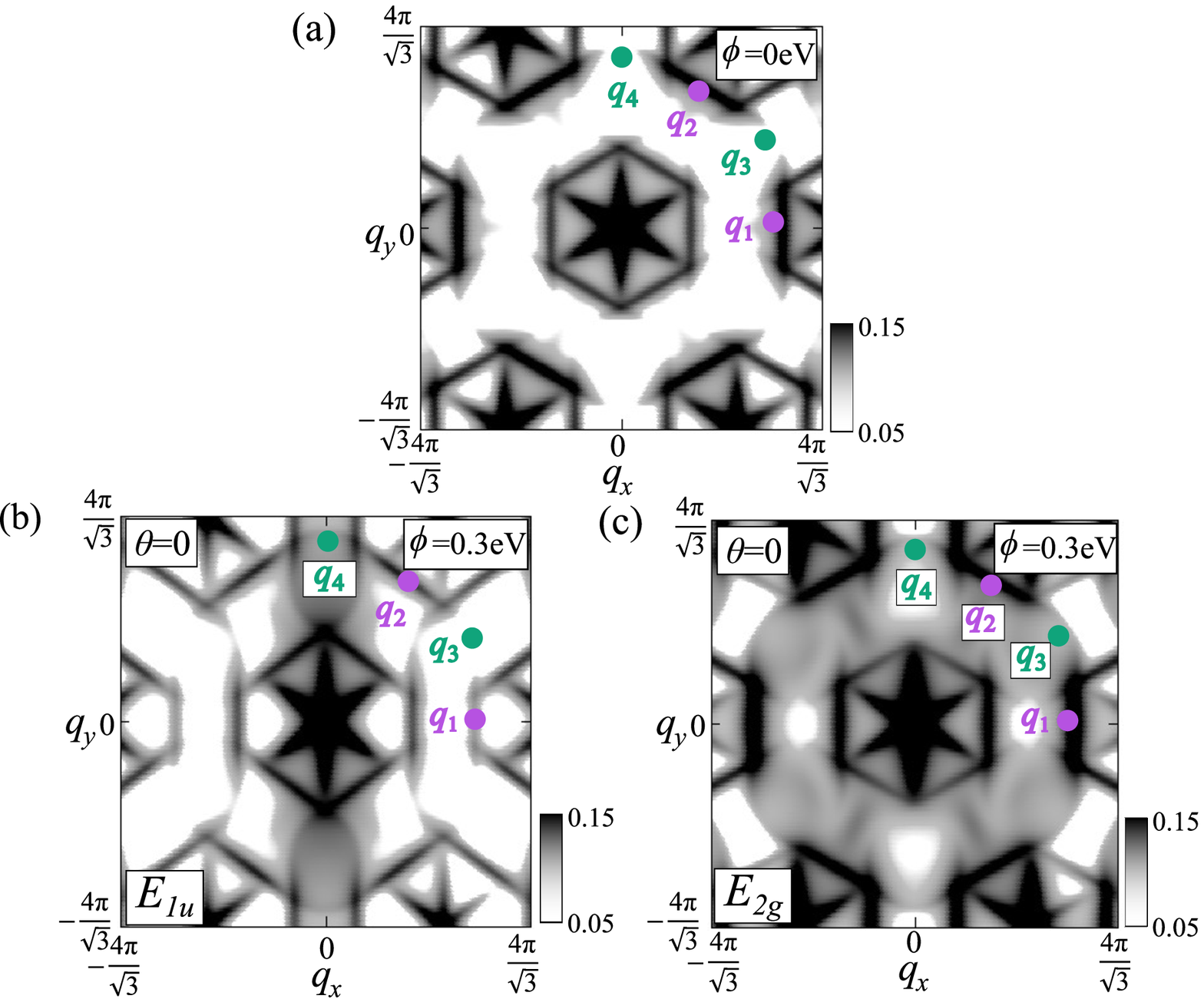}
\caption{
Normalized quasiparticle interference (QPI) signal
(a) at $\phi=0$,
(b) in the $E_{1u}$ bond order (BO) state, and
(c) in the $E_{2g}$ BO state at $\phi=0.3$eV.
}
\label{fig:QPI}
\end{figure}

Figure \ref{fig:QPI} shows the $\q$ dependence of the 
QPI signal due to the intra-$xy$-orbital FS scattering,
in the cases of (a) without BO ($\phi=0$),
(b) $E_{1u}$ BO ($\phi=0.3$eV), and
(c) $E_{2g}$ BO ($\phi=0.3$eV).
In the case of (b) the $E_{1u}$ BO,
the QPI signal becomes drastically anisotropic,
and the signal is strongly enlarged at $\q=\q_4$.
This result is consistent with the reports of the STM measurements 
in Refs. \cite{arXiv:2211.16477,arXiv:2211.12264}.
Note that $|\q_4|$ in this paper corresponds to $|\q_3|$ ($\q_4$)
in Ref. \cite{arXiv:2211.16477} (Ref. \cite{arXiv:2211.12264}).
In the case of (b) the $E_{2g}$ BO, in contrast,
the anisotropy of the QPI signal is almost isotropic,
which is inconsistent with experimental reports
\cite{arXiv:2211.16477,arXiv:2211.12264}.
Thus, it is concluded that the nematicity in CsTi$_3$Bi$_5$
originates from the odd-parity BO with $E_{1u}$ symmetry.

We verified that the large QPI anisotropy in the $E_{1u}$ state 
originates from the intersublattice component 
of the QP spectrum $\rho_{l,m}^\phi(\k,E)$ with $l \ne m$ 
in the JDOS in Eq. (\ref{eqn:QPI}).
The anisotropy of the QPI signal due to the intersublattice scattering
$\Delta I^\phi_{\rm inter}(\q_i) \equiv I^\phi_{\rm inter}(\q_i)-I^{\phi=0}_{\rm inter}(\q_i)$, 
is proportional to $\Delta G_{l,m}^\phi(\k_i)\Delta G_{m,l}^\phi(-\k_i)$,
where $\Delta G_{l,m}^\phi(\k) \equiv G_{l,m}^\phi(\k)-G_{l,m}^0(\k) \sim \phi G_{l,l}^0(k) f^{l,m}(\k)G_{m,m}^0(\k)$.
Therefore, in the case of the $E_{1u}$ BO,
$\Delta I^\phi_{\rm inter}(\q_i)\sim \phi^2 f^{l,m}(\k_i)f^{m,l}(-\k_i)=-\phi^2|f^{m,l}(\k_i)|^2$.
In the case of the $E_{2g}$ BO,
$\Delta I^\phi_{\rm inter}(\q_i)\sim \phi g^{l,m}(\k_i)g^{m,l}(-\k_i)=\phi|g^{m,l}(\k_i)|^2$.
The different sign of $\Delta I^\phi_{\rm inter}$ between odd and even-parity states gives rise to a qualitative difference in the QPI anisotropy.


Finally, we discuss why the FS deformation ($\Delta k_F$)
due to the $E_{1u}$ order ($\phi$) is proportional to $\phi^2$
based on the GL free-energy theory.
From the symmetry argument,
the third-order GL free energy with respect to 
the $E_{1u}$ order $\phi(\cos\theta,\sin\theta)$
and the $E_{2g}$ order $\eta(\cos\theta',\sin\theta')$
is given as $F'=-c\phi^2\eta\cos(2\theta-\theta'-\pi/2)$,
where $c$ is the GL coefficient.
Thus, for a fixed $\phi$,
the total GL free energy for $\eta$ up to the fourth order is
$F= a\eta^2+\frac{b}{2}\eta^4- (c\phi^2)\eta$ when $2\theta=\theta'$.
When $a>0$ (i.e., $T>T_{0}^{E_{2g}}$),
the secondary (or passive) $E_{2g}$ order is obtained as $\eta=(c\phi^2)/a$.
Because $\Delta k_F$ is proportional to the $E_{2g}$ order parameter,
we find $\Delta k_F\propto\eta\propto\phi^2 \ (\propto T_{0}^{E_{1u}}-T)$ 
when $T<T_{0}^{E_{1u}}$.

To summarize, the $E_{1u}$ BO gives the prominent nematicity
observed in CsTi$_3$Bi$_5$
\cite{arXiv:2211.16477,arXiv:2211.12264}.
The realized QPI signal anisotropy and the
FS deformation $\Delta k_F$ for $\phi\sim0.2$eV 
are comparable with the experimental reports
\cite{arXiv:2211.16477,arXiv:2211.12264}.
(Experimentally, $\Delta k_F/G\sim0.01$, where $G=4\pi/\sqrt{3}$ 
is the reciprocal lattice constant \cite{arXiv:2211.16477}.)
It is noteworthy that $\phi\sim0.2$eV is obtained by
the full DW equation analysis performed in Appendix D.
Importantly, $\Delta k_F \propto \phi^2$ in the odd-parity BO state,
while $\Delta k_F \propto \phi$ in the even-parity BO state.
This difference will be useful to determine the symmetry of the BO state.
We stress that the finite NLH effect is a crucial 
evidence for the odd-parity BO.
This is an important future issue in Ti-based kagome metals.


\subsection{APPENDIX H: RELATION TO $2\times2$ BO IN V-BASED KAGOME METALS\vspace{-1em}} 

We briefly discuss the relation between 
the present $E_{1u}$ BO and $3Q$ BO in AV$_3$Sb$_5$.
The quantum interference mechanism gives rise to the 
nearest site BO in both the Ti-based kagome metal ($xy$-orbital model) 
and the V-based one ($xz$-orbital model), while their wave vectors are different.
In the latter, since each VHS point consists of one sublattice
(sublattice interference),
the nearest site BO is given by the inter-VHS process with a finite wavevector. 
In contrast, in the former, $\k\approx \k_{\rm CA}$ consisting of C and A sublattices is important for the C-A site BO at $\q={\bm0}$.
Thus, the difference in the BO wavevector is naturally understood.
The $E_{1u}$ form factor $f^{l,m}(\k)$ 
given in Fig. \ref{fig:fig4}(a) [$(l,m)=({\rm B,C}),({\rm C,A})$]
is similar to the form factor of the V-based kagome metal
$g_{\q_i}^{l,m}(\k-\q_i/2)$, where $\q_i$ is the CDW wave vector
\cite{Tazai-kagome}.
Both BOs are caused by the paramagnon interference mechanism.



\begin{thebibliography}{99}

\bibitem{kagome-exp1}
B. R. Ortiz, L. C. Gomes, J. R. Morey, M. Winiarski, M. Bordelon, J. S. Mangum, I. W. H. Oswald, J. A. Rodriguez-Rivera, J. R. Neilson, S. D. Wilson, et al.,
{\it New kagome prototype materials: Discovery of KV$_{3}$Sb$_{5}$,RbV$_{3}$Sb$_{5}$, and Cs$V_{3}$Sb$_{5}$},
Phys. Rev. Materials {\bf 3}, 094407 (2019).


\bibitem{kagome-exp2}
B. R. Ortiz, S. M. L. Teicher, Y. Hu, J. L. Zuo, P. M. Sarte, E. C. Schueller, A. M. M. Abeykoon, M. J. Krogstad, S. Rosenkranz, et al.,
{\it CsV$_{3}$Sb$_{5}$: A $Z_{2}$ Topological Kagome Metal with a Superconducting Ground State},
Phys. Rev. Lett. {\bf 125}, 247002 (2020).


\bibitem{STM1}
Y.-X. Jiang, J.-X. Yin, M. M. Denner, N. Shumiya, B. R. Ortiz, G. Xu, Z. Guguchia, J. He, M. S. Hossain, X. Liu, J. Ruff, L. Kautzsch, S. S. Zhang, G. Chang, I. Belopolski, Q. Zhang, T. A. Cochran, D. Multer, M. Litskevich, Z.-J. Cheng, X. P. Yang, Z. Wang, R. Thomale, T. Neupert, S. D. Wilson, and M. Z. Hasan,
{\it Unconventional chiral charge order in kagome superconductor KV$_3$Sb$_5$},
Nat. Mater. {\bf 20}, 1353--1357 (2021).


\bibitem{STM2}
H. Li, H. Zhao, B. R. Ortiz, T. Park, M. Ye, L. Balents, Z. Wang, S. D. Wilson, and I. Zeljkovic,
{\it Rotation symmetry breaking in the normal state of a kagome superconductor KV$_3$Sb$_5$},
Nat. Phys. {\bf 18}, 265--270 (2022).

\bibitem{muSR3-Cs}
L. Yu, C. Wang, Y. Zhang, M. Sander, S. Ni, Z. Lu, S. Ma, Z. Wang, Z. Zhao, H. Chen, K. Jiang, Y. Zhang, H. Yang, F. Zhou, X. Dong, S. L. Johnson, M. J. Graf, J. Hu, H.-J. Gao, and Z. Zhao,
{\it Evidence of a hidden flux phase in the topological kagome metal CsV$_3$Sb$_5$},
arXiv:2107.10714 (avalable at https://arxiv.org/abs/2107.10714).


\bibitem{muSR2-K}
C. Mielke, D. Das, J.-X. Yin, H. Liu, R. Gupta, Y.-X. Jiang, M. Medarde, X. Wu, H. C. Lei, J. Chang, P. Dai, Q. Si, H. Miao, R. Thomale, T. Neupert, Y. Shi, R. Khasanov, M. Z. Hasan, H. Luetkens, and Z. Guguchia,
{\it Time-reversal symmetry-breaking charge order in a kagome superconductor},
Nature {\bf 602}, 245--250 (2022).

\bibitem{muSR4-Cs}
R. Khasanov, D. Das, R. Gupta, C. Mielke, M. Elender, Q. Yin, Z. Tu, C. Gong, H. Lei, E. T. Ritz, R. M. Fernandes, T. Birol, Z. Guguchia, and H. Luetkens,
{\it Time-reversal symmetry broken by charge order in ${\mathrm{CsV}}_{3}{\mathrm{Sb}}_{5}$},
Phys. Rev. Research {\bf 4}, 023244 (2022).

\bibitem{muSR5-Rb}
Z. Guguchia, C. Mielke, D. Das, R. Gupta, J.-X. Yin, H. Liu, Q. Yin, M. H. Christensen, Z. Tu, C. Gong, N. Shumiya, M. S. Hossain, T. Gamsakhurdashvili, M. Elender, P. Dai, A. Amato, Y. Shi, H. C. Lei, R. M. Fernandes, M. Z. Hasan, H. Luetkens, and R. Khasanov,
{\it Tunable unconventional kagome superconductivity in charge ordered RbV$_3$Sb$_5$ and KV$_3$Sb$_5$},
Nat. Commun. {\bf 14}, 153 (2023).

\bibitem{eMChA}
C. Guo, C. Putzke, S. Konyzheva, X. Huang, M. Gutierrez-Amigo, I. Errea, D. Chen, M. G. Vergniory, C. Felser, M. H. Fischer, T. Neupert, and P. J. W. Moll,
{\it Switchable chiral transport in charge-ordered kagome metal CsV$_3$Sb$_5$},
Nature {\bf 611}, 461--466 (2022).

\bibitem{elastoresistance-kagome}
L. Nie, K. Sun, W. Ma, D. Song, L. Zheng, Z. Liang, P. Wu, F. Yu, J. Li, M. Shan, D. Zhao, S. Li, B. Kang, Z. Wu, Y. Zhou, K. Liu, Z. Xiang, J. Ying, Z. Wang, T. Wu, and X. Chen,
{\it Charge-density-wave-driven electronic nematicity in a kagome superconductor},
Nature {\bf 604}, 59--64 (2022).

\bibitem{birefringence-kagome}
Y. Xu, Z. Ni, Y. Liu, B. R. Ortiz, Q. Deng, S. D. Wilson, B. Yan, L. Balents, and L. Wu,
{\it Three-state nematicity and magneto-optical Kerr effect in the charge density waves in kagome superconductors},
Nat. Phys. {\bf 18}, 1470--1475 (2022).

\bibitem{Roppongi}
M. Roppongi, K. Ishihara, Y. Tanaka, K. Ogawa, K. Okada, S. Liu, K. Mukasa, Y. Mizukami, Y. Uwatoko, R. Grasset, M. Konczykowski, B. R. Ortiz, S. D. Wilson, K. Hashimoto, and T. Shibauchi,
{\it Bulk evidence of anisotropic $s$-wave pairing with no sign change in the kagome superconductor CsV$_3$Sb$_5$},
arXiv:2206.02580 (avalable at https://arxiv.org/abs/2206.02580).

\bibitem{SC2}
W. Zhang, X. Liu, L. Wang, C. W. Tsang, Z. Wang, S. T. Lam, W. Wang, J. Xie, X. Zhou, Y. Zhao, S. Wang, J. Tallon, K. T. Lai, and S. K. Goh,
{\it Nodeless superconductivity in kagome metal CsV$_{3}$Sb$_{5}$ with and without time reversal symmetry breaking},
arXiv:2301.07374 (avalable at https://arxiv.org/abs/2301.07374).

\bibitem{arXiv:2304.06436}
Z. Guguchia, D.J. Gawryluk, Soohyeon Shin, Z. Hao, C. Mielke III, D. Das, I. Plokhikh, L. Liborio, K. Shenton, Y. Hu, V. Sazgari, M. Medarde, H. Deng, Y. Cai, C. Chen, Y. Jiang, A. Amato, M. Shi, M.Z. Hasan, J.-X. Yin, R. Khasanov, E. Pomjakushina, and H. Luetkens,
{\it Hidden magnetism uncovered in charge ordered bilayer kagome material ScV$_6$Sn$_6$},
arXiv:2304.06436


\bibitem{Thomale2021}
X. Wu, T. Schwemmer, T. M\"uller, A. Consiglio, G. Sangiovanni, D. Di Sante, Y. Iqbal, W. Hanke, A. P. Schnyder, M. M. Denner, M. H. Fischer, T. Neupert, and R. Thomale,
{\it Nature of Unconventional Pairing in the Kagome Superconductors $A{\mathrm{V}}_{3}{\mathrm{Sb}}_{5}$ ($A=\mathrm{K},\mathrm{Rb},\mathrm{Cs}$)},
Phys. Rev. Lett. {\bf 127}, 177001 (2021).


\bibitem{Neupert2021}
M. M. Denner, R. Thomale, and T. Neupert,
{\it Analysis of Charge Order in the Kagome Metal $A{\mathrm{V}}_{3}{\mathrm{Sb}}_{5}$ ($A=\mathrm{K},\mathrm{Rb},\mathrm{Cs}$)},
Phys. Rev. Lett. {\bf 127}, 217601 (2021).

\bibitem{Balents2021}
T. Park, M. Ye, and L. Balents,
{\it Electronic instabilities of kagome metals: Saddle points and Landau theory},
Phys. Rev. B {\bf 104}, 035142 (2021).

\bibitem{Tazai-kagome}
R. Tazai, Y. Yamakawa, S. Onari, and H. Kontani,
{\it Mechanism of exotic density-wave and beyond-Migdal unconventional superconductivity in kagome metal AV$_{3}$Sb$_{5}$ (A = K, Rb, Cs)},
Sci. Adv. {\bf 8}, eabl4108 (2022).

\bibitem{Tazai-kagome2}
R. Tazai, Y. Yamakawa and H. Kontani,
{\it Charge-loop current order and Z3 nematicity mediated by bond-order fluctuations in kagome metals},
Nat. Commun. {\bf 14}, 7845 (2023).

\bibitem{Tazai-kagome-GL}
R. Tazai, Y. Yamakawa and H. Kontani,
{\it Drastic magnetic-field-induced chiral current order and emergent current-bond-field interplay in kagome metals},
Proceedings of the National Academy of Sciences (PNAS) {\bf 121}, e2303476121 (2024).

\bibitem{Fernandes-GL}
M. H. Christensen, T. Biro, B. M. Andersen, and R. M. Fernandes,
{\it Loop currents in AV3Sb5 kagome metals: Multipolar and toroidal magnetic orders},
Phys. Rev. B {\bf 106}, 144504 (2022)

\bibitem{Thomale-GL}
F. Grandi, A. Consiglio, M. A. Sentef, R. Thomale, and D. M. Kennes,
{\it Theory of nematic charge orders in kagome metals},
Phys. Rev. B {\bf 107}, 155131 (2023)

\bibitem{Nat-g-ology}
H. D. Scammell, J. Ingham, T. Li, and O. P. Sushkov,
{\it Chiral excitonic order from twofold van Hove singularities in kagome metals},
Nat. Commun. {\bf 14}, 605 (2023)

\bibitem{fRG for vHS}
W. S. Wang, Z. Z. Li, Y. Y Xiang, and Q. H. Wang,
{\it Competing electronic orders on kagome lattices at van Hove filling}
Phys. Rev. B {\bf 87}, 115135 (2013).

\bibitem{fRG for nem}
F. Grandi, M. A. Sentef, D. M. Kennes, and R. Thomale,
{\it Theories for charge-driven nematicity in kagome metals}
arXiv:2406.09910.

\bibitem{Ti-ARPES}
Y. Hu, C. Le, Z. Zhao, J. Ma, N. C. Plumb, M. Radovic, A. P. Schnyder, X. Wu, H. Chen, X. Dong, J. Hu, H. Yang, H.-J. Gao, and M. Shi,
{\it Non-trivial band topology and orbital-selective electronic nematicity in a titanium-based kagome superconductor},
Nat. Phys. {\bf 19}, 1827 (2023).

\bibitem{CsTiBi5-oscillation}
Z. Rehfuss, C. Broyles, D. Graf, Y. Li, H. Tan, Z. Zhao, J. Liu, Y. Zhang, X. Dong, H. Yang, H. Gao, B. Yan, and S. Ran,
{\it Quantum Oscillations in kagome metals CsTi$_3$Bi$_5$ and RbTi$_3$Bi$_5$},
arXiv:2401.13628.

\bibitem{CsTiBi5-first}
H. Yang, Z. Zhao, X.-W. Yi, J. Liu, J.-Y. You, Y. Zhang, H. Guo, X. Lin, C. Shen, H. Chen, X. Dong, G. Su, and H.-J. Gao,
{\it Titanium-based kagome superconductor CsTi$_3$Bi$_5$ and topological states},
arXiv:2209.03840.

\bibitem{arXiv:2209.11656}
D. Werhahn, B. R. Ortiz, A. K. Hay, S. D. Wilson, R. Seshadri, and D. Johrendt,
{\it The kagome metals RbTi$_3$Bi$_5$ and CsTi$_3$Bi$_5$},
Z. Naturforsch. {\bf 77b}, 757 (2022).

\bibitem{arXiv:2211.16477}
H. Li, S. Cheng, B. R. Ortiz, H. Tan, D. Werhahn, K. Zeng, D. Jorhendt, B. Yan, Z. Wang, S. D. Wilson, and I. Zeljkovic,
{\it Electronic nematicity in the absence of charge density waves in a new titanium-based kagome metal},
Nat. Phys. {\bf 19}, 1591 (2023).

\bibitem{arXiv:2211.12264}
H. Yang, Y. Ye, Z. Zhao, J. Liu, X.-W. Yi, Y. Zhang, J. Shi, J.-Y. You, Z. Huang, B. Wang, J. Wang, H. Guo, X. Lin, C. Shen, W. Zhou, H. Chen, X. Dong, G. Su, Z. Wang, H.-J. Gao,
{\it Superconductivity and orbital-selective nematic order in a new titanium-based kagome metal CsTi$_3$Bi$_5$},
arXiv:2211.12264

\bibitem{ARPES-theory}
C. Bigi, M. Durrnagel, L. Klebl, A. Consiglio, G. Pokharel, F. Bertran, P.L. Fevre, T. Jaouen, H. C. Tchouekem, P.Turban, A.D. Vita, J.A. Miwa, J.W. Wells, D. Oh, R. Comin, R. Thomale, I. Zeljkovic, B.R. Ortiz, S.D. Wilson, G. Sangiovanni, F. Mazzola, D.D. Sante,
{\it Pomeranchuk instability from electronic correlations in CsTi$_3$Bi$_5$ kagome metal}
arXiv:2410.22929

\bibitem{CsTiBi5-ARPES}
Y. Wang {\it et al.},
{\it Flat Band and $Z_2$ Topology of Kagome Metal CsTi$_3$Bi$_5$},
Chin. Phys. Lett. {\bf 40}, 037102 (2023).

\bibitem{CsTiBi5-transport}
X. Chen, X. Liu, W. Xia, X. Mi, L. Zhong, K. Yang, L. Zhang, Y. Gan, Y. Liu, G. Wang, A. Wang, Y. Chai, J. Shen, X. Yang, Y. Guo, and M. He,
{\it Electrical and thermal transport properties of the kagome metals ATi$_3$Bi$_5$ (A=Rb,Cs)},
Phys. Rev. B {\bf 107}, 174510 (2023).

\bibitem{Onari-SCVC}
S. Onari and H. Kontani,
{\it Self-consistent Vertex Correction Analysis for Iron-based Superconductors: Mechanism of Coulomb Interaction-Driven Orbital Fluctuations},
Phys. Rev. Lett. {\bf 109}, 137001 (2012).

\bibitem{Yamakawa-FeSe}
Y. Yamakawa, S. Onari, and H. Kontani,
{\it Nematicity and Magnetism in FeSe and Other Families of Fe-Based Superconductors},
Phys. Rev. X {\bf 6}, 021032 (2016).

\bibitem{Onari-TBG}
S. Onari and H. Kontani,
{\it SU(4) Valley+Spin Fluctuation Interference Mechanism for Nematic Order in Magic-Angle Twisted Bilayer Graphene: The Impact of Vertex Corrections},
Phys. Rev. Lett. {\bf 128}, 066401 (2022).

\bibitem{Tsuchiizu1}
M. Tsuchiizu, Y. Ohno, S. Onari, and H. Kontani,
{\it Orbital Nematic Instability in the Two-Orbital Hubbard Model: Renormalization-Group + Constrained RPA Analysis},
Phys. Rev. Lett. {\bf 111}, 057003 (2013).

\bibitem{fRG-sin-orb}
R. Tazai, Y. Yamakawa, M. Tsuchiizu, H. Kontani,
{\it Functional renormalization group study of orbital fluctuation mediated superconductivity: Impact of the electron-boson coupling vertex corrections},
Phys. Rev. B {\bf 94}, 115155 (2016)

\bibitem{Tsuchiizu4}
M. Tsuchiizu, K. Kawaguchi, Y. Yamakawa, and H. Kontani,
{\it Multistage electronic nematic transitions in cuprate superconductors: A functional-renormalization-group analysis},
Phys. Rev. B {\bf 97}, 165131 (2018).

\bibitem{fRG-BEDT}
R. Tazai, Y. Yamakawa, M. Tsuchiizu, H. Kontani,
{\it Prediction of pseudogap formation due to $d$-wave bond-order in organic superconductor $\kappa$-(BEDT-TTF)$_2$X},
Phys. Rev. Research {\bf 3}, L022014 (2021)

\bibitem{Tazai-rev2021}
R. Tazai, Y. Yamakawa, M. Tsuchiizu, and H. Kontani,
{\it d- and p-wave Quantum Liquid Crystal Orders in Cuprate Superconductors, $\kappa$-(BEDT-TTF)$_2$X, and Coupled Chain Hubbard Models: Functional-renormalization-group Analysis},
J. Phys. Soc. Jpn. {\bf 90}, 111012 (2021)

\bibitem{Chubukov-PRX2016}
A. V. Chubukov, M. Khodas, and R. M. Fernandes,
{\it Magnetism, Superconductivity, and Spontaneous Orbital Order in Iron-Based Superconductors: Which Comes First and Why?},
Phys. Rev. X {\bf 6}, 041045 (2016).

\bibitem{Fernandes-rev2018}
R. M. Fernandes, P. P. Orth, and J. Schmalian,
{\it Intertwined Vestigial Order in Quantum Materials: Nematicity and Beyond},
Annu. Rev. Condens. Matter Phys. {\bf 10}, 133 (2019).

\bibitem{Kontani-AdvPhys}
H. Kontani, R. Tazai, Y. Yamakawa, and S. Onari,
{\it Unconventional density waves and superconductivities in Fe-based superconductors and other strongly correlated electron systems},
Adv. Phys. {\bf 70}, 355 (2021).

\bibitem{Davis-rev2013}
J. C. S. Davis and D.-H. Lee,
{\it Concepts relating magnetic interactions, intertwined electronic orders, and strongly correlated superconductivity},
Proc. Natl. Acad. Sci. U.S.A. {\bf 110}, 17623 (2013).

\bibitem{Shibauchi-FeSeTe}
K. Mukasa, K. Ishida, S. Imajo, M. Qiu, M. Saito, K. Matsuura, Y. Sugimura, S. Liu, Y. Uezono, T. Otsuka, M. Culo, S. Kasahara, Y. Matsuda, N. E. Hussey, T. Watanabe, K. Kindo, and T. Shibauchi,
{\it Enhanced Superconducting Pairing Strength near a Pure Nematic Quantum Critical Point},
Phys. Rev. X {\bf 13}, 011032 (2023).

\bibitem{Bickers2}
N. E. Bickers, D. J. Scalapino, S. R. White,
Phys. Rev. Lett. {\bf 62}, 961 (1989). 

\bibitem{Kontani-RH}
H. Kontani, K. Kanki, and K. Ueda,
Phys. Rev. B {\bf 59}, 14723 (1999).

\bibitem{Kontani-rev1}
H. Kontani,
Rep. Prog. Phys. {\bf 71}, 026501 (2008).

\bibitem{Tazai-LW}
R. Tazai, S. Matsubara, Y. Yamakawa, S. Onari, and H. Kontani,
{\it Rigorous formalism for unconventional symmetry breaking in Fermi liquid theory and its application to nematicity in FeSe},
Phys. Rev. B {\bf 107}, 035137 (2023).


\bibitem{Cd2Re2O7-1}
V. Kozii and L. Fu, 
{\it Odd-Parity Superconductivity in the Vicinity of Inversion Symmetry Breaking in Spin-Orbit-Coupled Systems},
Phys. Rev. Lett. {\bf 115}, 207002 (2015).

\bibitem{Cd2Re2O7-2}
S. Hayami, Y. Yanagi, H. Kusunose, and Y. Motome,
{\it Electric Toroidal Quadrupoles in the Spin-Orbit-Coupled Metal Cd$_2$Re$_2$O$_7$},
Phys. Rev. Lett. {\bf 122}, 147602 (2019).

\bibitem{Cd2Re2O7-3}
H. T. Hirose, T. Terashima, D. Hirai, Y. Matsubayashi, N. Kikugawa, D. Graf, K. Sugii, S. Sugiura, Z. Hiroi, and S. Uji,
{\it Electronic states of metallic electric toroidal quadrupole order in Cd$_2$Re$_2$O$_7$ determined by combining quantum oscillations and electronic structure calculations},
Phys. Rev. B {\bf 105}, 035116 (2022).

\bibitem{SHG-Cd2Re2O7}
J. W. Harter, Z. Y. Zhao, J.-Q. Yan, D. G. Mandrus, and D. Hsieh,
{\it A parity-breaking electronic nematic phase transition in the spin-orbit coupled metal Cd$_2$Re$_2$O$_7$},
Science {\bf 356}, 6335 (2017).

\bibitem{Onari-b2g}
S. Onari and H. Kontani,
{\it Origin of diverse nematic orders in Fe-based superconductors: $45^\circ$ rotated nematicity in AFe$_2$As$_2$ (A=Cs,Rb)}, 
Phys. Rev. B {\bf 100}, 020507(R) (2019).

\bibitem{polar-metal}
Y.-W. Fang and H. Chen,
{\it Design of a multifunctional polar metal via first-principles high-throughput structure screening},
Communications Materials {\bf 1}, 1 (2020).

\bibitem{Nem-R-1}
Y. Mizukami, O. Tanaka, K. Ishida, M. Tsujii, T. Mitsui, S. Kitao, M. Kurokuzu, M. Seto, S. Ishida, A. Iyo, H. Eisaki, K. Hashimoto, and T. Shibauchi
{\it Thermodynamic Signatures of Diagonal Nematicity in RbFe$_2$As$_2$ Superconductor},
arXiv: 2108.13081

\bibitem{Nem-R-2}
J.P. Sun, K. Matsuura, G.Z. Ye, Y. Mizukami, M. Shimozawa, K. Matsubayashi, M. Yamashita, T. Watashige, S. Kasahara, Y. Matsuda, J.-Q. Yan, B.C. Sales, Y. Uwatoko, J.-G. Cheng, and T. Shibauchi,
{\it Dome-shaped magnetic order competing with high-temperature superconductivity at high pressures in FeSe},
Nat. Commun. {\bf 7}, 12146 (2016).



\bibitem{JDOS-QPI-1}
J.E. Hoffman, K. McElroy, D.-H. Lee, K.M. Lang, H. Eisaki, and J.C. Davis,
{\it Imaging Quasiparticle Interference in Bi$_2$Sr$_2$CaCu$_2$O$_\{8+\delta\}$},
Science {\bf 297}, 1148 (2002).

\bibitem{JDOS-QPI-2}
K. McElroy, G.-H. Gweon, S.Y. Zhou, J. Graf, S. Uchida, H. Eisaki, H. Takagi, T. Sasagawa, D.-H. Lee, and A. Lanzara,
{\it Elastic Scattering Susceptibility of the High Temperature Superconductor Bi$_2$Sr$_2$CaCu$_2$O$_\{8+\delta\}$: A Comparison between Real and Momentum Space Photoemission Spectroscopies},
Phys. Rev. Lett. {\bf 96}, 067005 (2006).

\bibitem{JDOS-QPI-3}
J.W. Alldredge, K. Fujita, H. Eisaki, S. Uchida, and K. McElroy,
{\it Three-component electronic structure of the cuprates derived from spectroscopic-imaging scanning tunneling microscopy},
Phys. Rev. B {\bf 85}, 174501 (2012).


\bibitem{NLH-LFu}
I. Sodemann and L. Fu, 
{\it Quantum nonlinear Hall effect induced by Berry curvature dipole in time-reversal invariant materials},
Phys. Rev. Lett. {\bf 115}, 216806 (2015).

\bibitem{NLH-WTe}
Q. Ma {\it et al}., 
{\it Observation of the nonlinear Hall effect under time-reversal-symmetric conditions},
Nature {\bf 565}, 337 (2019).

\bibitem{NLH-TBG}
M. Huang, Z. Wu, X. Zhang, X. Feng, Z. Zhou, S. Wang, Y. Chen, C. Cheng, K. Sun, Z. Y. Meng, and N. Wang, 
{\it Intrinsic Nonlinear Hall Effect and Gate-Switchable Berry Curvature Sliding in Twisted Bilayer Graphene},
Phys. Rev. Lett. {\bf 131}, 066301 (2023).

\bibitem{NLH-review}
Z. Z. Du, H.-Z. Lu and X. C. Xie,
{\it Nonlinear Hall effects},
Nature Reviews Physics {\bf 3}, 744 (2021).

\bibitem{Morimoto-review}
J. Orenstein, J.E. Moore, T. Morimoto, D.H. Torchinsky, J.W. Harter, and D. Hsieh,
{\it Topology and Symmetry of Quantum Materials via Nonlinear Optical Responses},
Annu. Rev. Cond. Mat. Phys. {\bf 12}, 247 (2021).

\bibitem{Yanase-JPSJ}
Y. Yanase, 
{\it Magneto-Electric Effect in Three-Dimensional Coupled Zigzag Chains},
J. Phys. Soc. Jpn. {\bf 83}, 014703 (2014).

\bibitem{Kontani-sLC}
H. Kontani, Y. Yamakawa, R. Tazai, S. Onari,
{\it Odd-parity spin-loop-current order mediated by transverse spin fluctuations in cuprates and related electron systems},
Phys. Rev. Research {\bf 3},013127 (2021)

\end{thebibliography}
\end{document}